\documentclass[usenatbib]{mn2e}
\usepackage{graphicx}
\usepackage{amssymb}
\usepackage{amsmath}
\usepackage{color}

\usepackage{graphicx}
\newcommand{\bj}{\bmath{j}}

\newcommand{\bb}{\bmath{b}}

\newcommand{\bv}{\bmath{v}}
\newcommand{\bu}{\bmath{u}}

\newcommand{\bB}{\bmath{B}}
\newcommand{\bE}{\bmath{E}}

\newcommand{\cQ}{{\cal Q}}
\newcommand{\cF}{{\cal F}}
\newcommand{\cS}{{\cal S}}
\newcommand{\cP}{{\cal P}}
\newcommand{\cL}{{\cal L}}
\newcommand{\cD}{{\cal D}}
\newcommand{\cM}{{\cal M}}
\newcommand{\cE}{{\cal E}}

\newcommand{\cDF}{{\cal D\!F}}

\newcommand{\etext}[1]{\quad\mbox{#1}\quad}
\newcommand{\spr}[2]{\bmath{#1} \!\cdot\! \bmath{#2}}

\newcommand{\oder}[2]{\frac{d #1}{d #2}}
\newcommand{\pder}[2]{\frac{\partial #1}{\partial #2}}
\newcommand{\Pd}[1]{\partial_{#1}}
\newcommand{\fracp}[2]{\left(\frac{#1}{#2}\right)}

\newcommand{\ort}[1]{ \bmath{i}_{#1} }

\newcommand{\sub}[1]{_{\mbox{\scriptsize #1}}}
\newcommand{\beq}{\begin{equation}}
\newcommand{\eeq}{\end{equation}}

\newcommand{\Km}{\mathcal K\sub{m}}
\newcommand{\Kf}{\mathcal K\sub{f}}
\newcommand{\Kq}{\mathcal K\sub{q}}
\newcommand{\sech}{\mbox{sech}}

\onecolumn

\title[Two-fluid RMHD]{A multi-dimensional numerical scheme for two-fluid Relativistic MHD}
\author[Barkov \& Komissarov]{Maxim Barkov$^{1,3}$\thanks{E-mail: bmv@mpi-hd.mpg.de (MB)},
Serguei S. Komissarov$^{2}$\thanks{E-mail: serguei@maths.leeds.ac.uk (SSK)},
Vitaly Korolev$^{4}$\thanks{E-mail: vitokorolev@gmail.com (VK)},
Andrey Zankovich$^{4}$\thanks{E-mail: zed81@list.ru (AZ)} \\
$^{1}$ Max-Planck-Institut f\"ur Kernphysik, Saupfercheckweg 1, 69117 Heidelberg, Germany \\
$^{2}$ Department of Applied Mathematics, The University of Leeds, Leeds, LS2 9GT, UK \\
$^{3}$ Space Research Institute RAS, 84/32 Profsoyuznaya Street, Moscow, 117997, Russia\\
$^{4}$ Volgograd State University, Volgograd 400062, Russia}

\begin{document}
\date{Received/Accepted}
\maketitle

\begin{abstract} 
The paper describes an explicit multi-dimensional numerical scheme for Special Relativistic 
Two-Fluid Magnetohydrodynamics of electron-positron plasma and a suit of test problems.   
The scheme utilizes Cartesian grid and the third order
WENO interpolation. The time integration is carried out using the third order TVD
method of Runge-Kutta type, thus ensuring overall third order accuracy on smooth
solutions. The magnetic field is kept near divergence-free by means of  
the method of generalized Lagrange multiplier. The test simulations, which include 
linear and non-linear continuous plasma waves, shock waves, strong explosions and 
the tearing instability, show that the scheme is sufficiently robust and confirm 
its accuracy.  

\end{abstract}
                                                                                          
                                                                                          
\section{Introduction}
\label{introduction}

It is now well recognized that magnetic fields play a very important role 
in many astrophysical phenomena and in particular in those involving relativistic 
outflows. The magnetic fields are likely to be involved in launching, powering and 
collimation of such outflows. 
The dynamics of relativistic magnetized plasma can be studied using diverse 
mathematical frameworks. The most developed one so far is the single fluid 
ideal relativistic Magnetohydrodynamics (RMHD). During the last decade, an impressive 
progress has been achieved in developing robust and efficient computer 
codes for integration of its equations. Only within a scope of proper review 
one can acknowledge the numerous contributions made by many different research 
groups and individuals. This framework is most suitable for studying large-scale 
macroscopic motions as it does not require to resolve the microphysical scales for 
numerical stability. One of its obvious disadvantages is that it allows 
only numerical dissipation.  While at shocks this is not of a problem, the  
dissipation observed in simulations at other locations may be questionable. 
This becomes more of an issue as ever growing body of evidence suggests 
the importance of magnetic dissipation associated with magnetic reconnection 
in dynamics of highly magnetized relativistic plasma.  The framework of resistive 
RMHD allows to incorporate this magnetic dissipation but the inevitably 
phenomenological nature of its Ohm's law puts constraints on its robustness. 
It is not yet clear how much progress can be achieved in this direction.

At the other extreme is the particle dynamics, describing the motion of 
individual charges in their collective electromagnetic field. The so-called 
Particle-in-Cell (PIC) numerical approach has been extremely productive in 
studying the micro-physics of relativistic collisionless shock waves and magnetic 
reconnection. The numerical stability considerations require PIC codes to resolve the 
scales of plasma oscillations. The accuracy considerations can be even more demanding, 
pushing towards the particle gyration scales. The accuracy of PIC simulations 
also depends on the number of macro-particles and rather weakly, with the 
numerical noise level being inversely proportional of the square root of the number.   
These features make this approach less suitable for studying macroscopic phenomena 
and only few examples of such simulations can be found in literature 
\citep[e.g.][]{ABSK95,SA02}.         

Somewhere in between there lays the multi-fluid approximation, where plasma is modeled 
as a collection of several inter-penetrating charged and neutral fluids, coupled 
via macroscopic electromagnetic field and inter-fluid friction. Undoubtedly, 
this approach is not as comprehensive in capturing the microphysics of collisionless 
plasma as the particle dynamics (and kinetics). However, it does this better than the single 
fluid MHD. In particular, it captures the collective interaction between positive and negative 
charges which results in plasma waves with frequencies above the plasma frequency. 
Compared to the resistive MHD, the generalized Ohm's law of the multi-fluid approximations
takes into account the finite inertia of charged particles. From the numerical prospective, 
the potential of this framework is not clear. By the present time, there has been only a very 
limited effort to explore it. \citet{ZHK09b,ZHK09a} used this approach for studying the 
relativistic magnetic reconnection, \citet{Amano13} for studying the termination 
shocks of pulsar winds, and \citet{KO09} tried to construct two-fluid models of 
steady-state pulsar magnetospheres. The ability, or the lack of it, to describe 
accurately the magnetic reconnection is probably the most important issue. 
\citet{ZHK09b} concluded that the inertial terms make a significant contribution to the 
reconnection electric field, even exceeding that of the friction term, which represents the 
resistivity. This hints that the inertial terms can play an important role in driving the 
magnetic reconnection. On the other hand, they have also found that the outcome strongly 
depends on the model of resistivity. Thus, the reconnection issue remains widely open.
Another issue is the required numerical resolution. \citet{ZHK09b,ZHK09a} set the grid scale 
to be comparable to the electron gyration radius. If such high resolution is indeed required
then it will be easier just to do PIC simulations. Only if much lower resolution is 
sufficient, the multi-fluid approach can be advantageous when it comes to macroscopic 
simulations.    
     
The aforesaid works on the relativistic multi-fluid numerical simulations 
neither described in details the design of their numerical schemes nor presented 
test simulations demonstrating their robustness and accuracy. 
\citet{ZHK09b,ZHK09a} state that they use a 
``modified Lax-Wendroff scheme'', which employs artificial viscosity, and comment that 
this is a ``bottleneck'' of their code.  In particular, they find that 
the contribution of the artificial viscosity to the reconnection electric field 
is comparable to other contributions. In this paper, we present a high-order accurate 
Godunov-type scheme for the two-fluid RMHD which does not involve such artificial terms. 
As a first step, we constrain ourselves with the Special Relativistic framework and 
the case of pure  electron-positron plasma. The scheme utilizes the third order WENO 
interpolation and the time integration is carried out using the third order TVD 
method of Runge-Kutta type, thus ensuring overall third order accuracy on smooth 
solutions. In order to keep the magnetic field divergence-free we employ 
the method of generalized Lagrange multiplier. The code is parallelized using MPI. 
We also present a suit of one-dimensional and multi-dimensional test problems and the 
results of our test simulations. The code is  parallelized using MPI. It can be 
provided by the authors on request.    

\section{3+1 Special Relativistic electron-positron two-fluid MHD}
\label{sec:TF-equations}

In the multi-fluid approximation, it is assumed that plasma consists 
of several interacting fluids, one per each kind of particles involved, 
and the electromagnetic field. Charged fluids interact with each other 
electromagnetically and via the so-called inter-fluid friction. The coupling between 
a fluid of electrically neutral particles and other fluids is only via the inter-fluid 
friction. Given such understanding, the system of multi-fluid approximation 
consists of the Maxwell equations and copies of the usual single fluid equations, 
one per each involved fluid, augmented with the force terms describing these interactions. 
The fluid equations may be derived via averaging of the corresponding kinetic 
equations \citep[e.g.][]{GP04}. 

In one form or another, the equations of relativistic two-fluid MHD were used by
 many researchers, going back to the work  by \citet{AP56}.  
Following \citet{ZHK09b} we adopt the 3+1 Special Relativistic equations originated from 
the covariant formulation by \citet{Gurovich86}.  In a global inertial frame with
time-independent coordinate grid, the equations of two-fluid RMHD include 
the continuity equations

\beq
   \Pd{t}(\rho_\pm \gamma_\pm) + \nabla_i(\rho_\pm u_\pm^i)=0 \, ,
\eeq
the energy equations  

\beq
   \Pd{t}(w_\pm\gamma_\pm^2 -p_\pm) + 
    \nabla_i(w_\pm\gamma_\pm u^i_\pm)  =  (j^i_\pm E_i) 
    \,\pm\, \varkappa\sub{f}c^2 n_+ n_- (\gamma_--\gamma_+) \, ,
\label{eq:en0}
\eeq
the momentum equations 

\beq
    \Pd{t}\left((w_\pm/c^2) \gamma_\pm u_\pm^s\right) +  
    \nabla_i((w_\pm/c^2) u^i_\pm u^s_\pm +g^{is}p_\pm) = 
     q_\pm E^s  +\frac{1}{c} e^{sik}j^\pm_i B_k 
    \,\pm\, \varkappa\sub{f} n_+ n_- (u^s_- - u^s_+) \, ,
\label{eq:mom0}
\eeq
and the Maxwell equations

\begin{equation}
   \nabla_i B^i=0 \, ,
\end{equation}

\begin{equation}
   \frac{1}{c}\Pd{t} B^s + e^{sik}\Pd{i}E_k = 0 \, ,
\end{equation}

\begin{equation}
   \nabla_i E^i = 4\pi (q_+ + q_-) \, , 
\end{equation}

\begin{equation}
   \frac{1}{c}\Pd{t} E^s - e^{sik}\Pd{i}B_k = -\frac{4\pi}{c}(j^s_+ + j^s_-) \, .
\end{equation}
We apply these to the case of electron-positron plasma, so 
$n_\pm$ is the proper number density of positrons and electrons 
(as measured in the rest frame of the positron and electron fluids respectively),
$\rho_\pm= m\sub{e}n_\pm$ is the proper mass density, $\gamma_\pm$ is
the Lorentz factor, $\bu_\pm=\gamma_\pm \bv_\pm$ are the spatial
components of 4-velocity, $w_\pm$ is the relativistic enthalpy,
$p_\pm$ is the isotropic thermodynamic pressure, $q_\pm =\pm e \gamma_\pm n_\pm$
is the electric charge density, $\bj_\pm = \pm e n_\pm \bu_\pm$ is the
electric current density, $g^{ij}$ is the metric tensor of space,
$e^{ijk}$ is the Levi-Civita tensor of space, and $c$ is the speed of
light.  The last two terms in Eqs.\ref{eq:en0} and \ref{eq:mom0} describe 
the internal friction between the two fluids. The 4-scalar $\varkappa\sub{f}$ is 
called the dynamic coefficient of this friction.
These equations have to be supplemented by equations of state (EOS), e.g. 
the polytropic EOS

\beq
w_\pm=n_\pm m_e c^2+\Gamma p_\pm /(\Gamma-1) \,,
\eeq
where $\Gamma$ is the ratio of specific heats.

By subtracting the momentum equations for the positron and electron fluids,
one obtains the so-called generalized Ohm law of the two-fluid MHD:

\beq
\Pd{t}\left( \sum\limits_{\pm} \pm (w_{\pm}/c^2) \gamma_{\pm}  u_{\pm}^s \right) +
\nabla_i \left(
   \sum\limits_{\pm} \pm ((w_{\pm}/c^2) u^i_{\pm} u^s_{\pm} +p_{\pm}g^{is})
  \right)   = 
     e \tilde{n} \left( E^s  +\frac{1}{c} e^{sik}v_i B_k \right) 
    - \frac{2\varkappa\sub{f}}{e} (n_+j^s_- + n_-j^s_+) \, ,
\label{Ohm_m0}
\eeq
In this equation, 

\beq
\tilde{n} = n_+\gamma_+ + n_-\gamma_-
\eeq
is the total number density of charged particles as measured in the 
laboratory frame and

\beq 
\bv=\frac{n_+\bu_+ + n_-\bu_-}{\tilde{n}} 
= \frac{\bj_+ - \bj_-}{e\tilde{n}}   
\eeq
is their average velocity in this frame. In general, the internal friction of a 
single fluid, not introduced in 
our model, leads to viscosity and the internal friction between two charged 
fluids to resistivity. In order to understand the properties of this resistivity,
consider the Ohm's law in the plasma frame (where $\bv=0$) and ignore the 
inertial terms (the terms in the left-hand side of the generalized Ohm's law). 
This gives us  
\beq
   \bE  = \frac{2\varkappa\sub{f}}{e^2\tilde{n}} (n_+\bj_- + n_-\bj_+) \, .
\label{Ohm_m1}
\eeq
Since $\bv=0$, we have $\bj_+=\bj_-=\bj/2$, where $\bj=\bj_+ +\bj_-$ is the total 
electric current, and hence obtain the usual Ohm's law   
\beq
   \bj = \sigma \bE\, ,
\label{Ohm_m2}
\eeq
with the scalar conductivity coefficient
\beq
   \sigma  = \frac{e^2}{\varkappa\sub{f}} \fracp{\tilde{n}}{n_+ + n_-} \, . 
\label{sigma}
\eeq
The corresponding resistivity coefficient is defined as $\eta=1/4\pi\sigma$. 

For the purpose of mathematical analysis and computations, it is best to deal 
with dimensionless equations.   
Denote as $B_0$ the characteristic value of magnetic (and electric)
field, as $T_0$ the characteristic time scale, as $L_0=cT_0$ the
characteristic length scale, and as $n_0$ the characteristic number density of particles.  
Then the characteristic scale for $\rho$ is $m\sub{e} n_0$ and the
characteristic scale for $w$ and $p$ is $m\sub{e} c^2 n_0$. The corresponding
dimensionless equations of the two-fluid electron-positron RMHD are

\beq 
\Pd{t}(n_\pm \gamma_\pm) + \nabla_i(n_\pm u_\pm^i)=0 \, ,
\label{continuity}
\eeq

\beq 
\Pd{t}(w_\pm\gamma_\pm^2 -p_\pm) + \nabla_i(w_\pm\gamma_\pm
u^i_\pm)= \pm \frac{1}{\Km} n_\pm(u^i_\pm E_i) \,\pm\, \frac{1}{\Kf}
n_+ n_- (\gamma_--\gamma_+) \, ,
\label{energy}
\eeq

\beq 
\Pd{t}\left(w_\pm \gamma_\pm u_\pm^s\right) + \nabla_i(w_\pm
u^i_\pm u^s_\pm +g^{is}p_\pm)= \pm \frac{1}{\Km} n_\pm \gamma_\pm
(E^s + e^{sik}v^\pm_i B_k) \,\pm\, \frac{1}{\Kf} n_+ n_- (u^s_- -
u^s_+) \, ,
\label{momentum}
\eeq

\begin{equation}
   \nabla_i B^i=0 \, ,
\label{Gauss_B}
\end{equation}

\begin{equation}
   \Pd{t} B^s + e^{sik}\Pd{i}E_k = 0 \, ,
\label{Faraday}
\end{equation}

\begin{equation}
\label{Gauss_E}
   \nabla_i E^i = \frac{1}{\Kq}(n_+\gamma_+ - n_-\gamma_-) \, ,
\end{equation}

\begin{equation}
   \Pd{t} E^s - e^{sik}\Pd{i}B_k = - \frac{1}{\Kq}(n_+u_+^s - n_-u^s_-) \, .
\label{Ampere}
\end{equation}
The three dimensionless parameters in these equations are
\beq 
\Kq = \frac{B_0}{4\pi e L_0 n_0} \,, \quad
\Km = \frac{m\sub{e}c^2}{e B_0 L_0} \,, \quad 
\Kf = \frac{m\sub{e}c}{\varkappa\sub{f} n_0 L_0} \,.
\label{Ks} 
\eeq 
The dimensionless polytropic EOS is  
\beq
w_\pm=n_\pm +\Gamma p_\pm /(\Gamma-1) \,.
\eeq

\begin{table}
\caption{Important plasma parameters in dimensional and dimensionless form. 
Here the magnetic field $\hat{B}$ is measured in the plasma rest frame, $w=w_++w_-$, 
$p=p_++p_-$ and $n=n_++n_-$.} 
\begin{center}
    \label{pp}
   \begin{tabular}{|c|c|c|}
   \hline
   \hline
       Name & Dimensional expression  & Dimensionless expression \\
   \hline
   \hline
     plasma frequency & $\Omega\sub{p}^2 = \frac{4\pi e^2 n^2 c^2}{w}$ & 
                        $\Omega\sub{p}^2 = \frac{1}{\Kq\Km}\fracp{n^2}{w}$\\
     skin depth       & $\lambda\sub{s} = \frac{c}{\Omega\sub{p}}$ & 
                        $\lambda\sub{s} = \fracp{1}{\Omega\sub{p}}$ \\
     magnetization $\beta$ & $ \beta=\frac{8\pi p}{\hat{B}^2} $ &
                             $\beta=\frac{\Km}{\Kq} \fracp{2 p}{\hat{B}^2}$ \\
     magnetization $\sigma\sub{m}$ & $\sigma\sub{m}=\frac{\hat{B}^2}{4\pi w}$ & 
                       $\sigma\sub{m}=\frac{\Kq}{\Km}\fracp{\hat{B}^2}{w} $ \\
     gyration radius  & $R\sub{g}=\frac{m\sub{e}c^2}{e\hat{B}} 
                         \left( \fracp{w-p}{\rho c^2}^2-1 \right)^{1/2}$ &  
                        $R\sub{g} = \frac{\Km}{\hat{B}} \left( \fracp{w-p}{n}^2-1 \right)^{1/2}$ \\ 
   \hline
   \hline
   \end{tabular}
\end{center}
\end{table}

We write the dimensionless two-fluid equations in the form which focuses on 
their mathematical structure. Although this somewhat hides the physical origin of the system,  
all the important plasma parameters can be easily recovered when this is needed.  
Some of them are listed in Table~\ref{pp}.   
The parameters $\Km$, $\Kf$, and $\Kq$ also allow simple physical interpretation. 
$\Km$ is the ratio of the particle rest mass energy
$m\sub{e}c^2$ to the potential drop $eB_0L_0$ in the electric field of
strength $E_0=B_0$ over the characteristic length scale of the
problem. Obviously, this is a measure of the magnetic field
strength. $\Kf$ is the ratio of the relaxation time scale
$T\sub{f}=m\sub{e}/n_0\varkappa\sub{f}$ in plasma of number density $n_0$
and the light-crossing time scale $T_0=L_0/c$. $\Kq$ is the inverse ratio of $n_0$ 
and the Goldreich-Julian number density of charged particles 

\beq 
    n\sub{GJ} = B_0/4\pi e L_0 \, .
\label{nGJ}
\eeq 

From the expressions for the plasma magnetizations $\beta$ and $\sigma\sub{m}$ 
given in Table~\ref{pp} it is 
easy to see how to setup two-fluid problems corresponding to single-fluid RMHD 
problems, e.g. the test problems of \citet{ssk-godun99}. One can
use the same dimensionless values for the total thermodynamic pressure
and density, but scale the electromagnetic field by the factor of $\sqrt{\Km/\Kq}$.

By summing together the energy equations Eqs.\ref{energy} and then  
utilizing the Maxwell equations one obtains the conservation 
law for the total energy, 

\beq 
\Pd{t} \left( \sum\limits_{\pm} ( w_{\pm}\gamma_{\pm}^2 -p_{\pm}) + 
            \frac{\Kq}{2\Km}(B^2+E^2) \right) 
+ \nabla_i \left(  \sum\limits_{\pm} w_{\pm}\gamma_{\pm} u^i_{\pm} + 
        \frac{\Kq}{\Km} e^{ijk}E_jB_k \right)= 0  \, .
\label{total-energy}
\eeq
Similarly, one finds the total momentum conservation law 
 
\beq 
\Pd{t}\left( \sum\limits_{\pm} w_{\pm} \gamma_{\pm}  u_{\pm}^s + 
   \frac{\Kq}{\Km} e^{sjk}E_jB_k   \right) + 
\nabla_i \left( \sum\limits_{\pm}(w_{\pm} u^i_{\pm} u^s_{\pm} +p_{\pm}g^{is}) +  
\frac{\Kq}{\Km}\left(-E^iE^s-B^iB^s + \frac{1}{2}(B^2+E^2)g^{is} \right)
\right)= 0\, .
\label{total-momentum}
\eeq

By subtracting the dimensionless momentum equations for the positron and electron fluids
one obtains the dimensionless generalized Ohm law of the two-fluid MHD: 

\beq
\Pd{t}\left( \sum\limits_{\pm} \pm w_{\pm} \gamma_{\pm}  u_{\pm}^s \right) + 
\nabla_i \left( 
   \sum\limits_{\pm} \pm (w_{\pm} u^i_{\pm} u^s_{} +p_{\pm}g^{is} )    
  \right)   =  
      \frac{1}{\Km}\tilde{n} (E^s+e^{sik}v_i B_k) + 
      \frac{2}{\Kf} n_+n_- (u^s_- - u^s_+)  \, .
\label{Ohm_m}
\eeq  
In this equation, $\tilde{n} = n_+\gamma_+ + n_-\gamma_-$ is the total number density 
of charged particles as measured in the laboratory frame and 
$v^i=(n_+\gamma_+v^i_+ + n_-\gamma_-v^i_-)/\tilde{n}$ is their average 
velocity in this frame. Subtracting the dimensionless energy equations, we 
obtain the time component of the generalized Ohm's law  
\beq
\Pd{t} \left( \sum\limits_{\pm} \pm ( w_{\pm}\gamma_{\pm}^2 -p_{\pm}) \right)
+ \nabla_i \left(  \sum\limits_{\pm} \pm(w_{\pm}\gamma_{\pm} u^i_{\pm})\right) 
   = \frac{1}{\Km} \tilde{n}(v^i E_i)  +  
\frac{2}{\Kf} n_+ n_- (\gamma_--\gamma_+) \, .
\label{Ohm_e}
\eeq

\section{Numerical method}
\label{num-method}

From the analysis of Sec.\ref{sec:TF-equations}, it follows that there is a 
significant freedom in the selection of mathematically equivalent closed system 
of equations to describe the dynamics of two-fluid plasma.  
However for numerical integration, it is often advantageous to select equations which 
do not involve source terms. In this case, one can utilize the ability  of  modern 
shock capturing schemes to preserve conserved quantities down to the computer rounding 
error, which is important for problems involving discontinuities.  
For this reason, the fluid equations utilized in our 
numerical scheme include Eqs.~\ref{continuity}, \ref{total-energy} and \ref{total-momentum}.    
The system is closed by including the generalized Ohm' law, Eqs.~\ref{Ohm_m} and \ref{Ohm_e}, 
and the so-called augmented Maxwell equations, which are described below. 

\subsection{Augmented Maxwell Equations} 

In order to keep the magnetic field divergence free and to ensure that
the Gauss law is satisfied as well, we use the so-called Generalized 
Lagrange Multiplier method \citep{Munz00,Dedner02,ssk-rrmhd}. 
To this end, the Maxwell equations are modified in such a 
way that all four of them become dynamic. Namely, 

\begin{equation}
   \Pd{t}\Phi + \nabla_i B^i=-\kappa\Phi \, ,
\label{GaussB_a}
\end{equation}

\begin{equation}
   \Pd{t} B^s + e^{sik}\Pd{i}E_k + \nabla^s {\Phi} = 0 \, ,
\label{Faraday_a}
\end{equation}

\begin{equation}
\label{GaussE_a}
   \Pd{t}\Psi + \nabla_i E^i = \frac{1}{\Kq}(n_+\gamma_+ + n_-\gamma_-)  -\kappa\Psi \, ,
\end{equation}

\begin{equation}
   \Pd{t} E^s - e^{sik}\Pd{i}B_k + \nabla^s \Psi  = - \frac{1}{\Kq}(n_+u_+^s - n_-u^s_-) \, ,
\label{Ampere_a} 
\end{equation}
where the scalars $\Phi$ and $\Psi$ are two additional dynamic variables 
and $\kappa$ is a positive constant, which determines the rate of decay of 
$\nabla_i B^i$ and $\nabla_i E^i-(1/\Kq)(n_+\gamma_+ + n_-\gamma_-)$.
Obviously, the resultant equations are of the same structure as the fluid equations and 
hence they can be discretized in exactly 
the same way. This property greatly simplifies computational algorithms and coding.     

\subsection{Finite-difference conservative scheme}
    
In our numerical scheme, we integrate 
Eqs.~\ref{continuity}, \ref{total-energy}, \ref{total-momentum}, \ref{Ohm_m}, \ref{Ohm_e}, 
and the augmented Maxwell equations \ref{GaussB_a}-\ref{Ampere_a}.
Taken together, they can be written as a single phase vector 
equation of the form
                                                                                          
\begin{equation}
  \pder{\cQ}{t} + \pder{\cF^{m}}{x^m}  = \cS \, ,
\label{CONS}
\end{equation}
where $\cQ$ is the vector of conserved variables, $\cF^m$ is the vector of 
fluxes in the $x^m$ direction and $\cS$ is the vector of source terms. When $\{x^m\}$ 
are Cartesian coordinates, and this is the only case we have coded so far, 
one can simply replace $\nabla_m$ with $\Pd{m}$ and hence read the components of 
$\cQ$, $\cF^m$ and $\cS$ directly from the aforesaid equations. Otherwise, these expressions 
will also involve components of the metric tensor.      

Introduce a uniform coordinate grid such that $x^m_i$ is the coordinate of the center of 
the $i$th cell, and $x^m_{i\pm 1/2}$ are the coordinates of its right and left interfaces. 
Replace, the continuous function $Q$ with a discrete function defined at the cell 
centers and the continuous functions $F^m$ with discrete functions defined at the 
centers of $x^m$ cell interfaces. 
Finally, replace $\Pd{m}{\cF^{m}}$  with the finite difference approximations:   

\beq
    \cDF_{i,j,k} =  
    \frac{\cF^{1}_{i+1/2,j,k} -\cF^{1}_{i-1/2,j,k} } {\Delta x^1} +
    \frac{\cF^{2}_{i,j+1/2,k} -\cF^{2}_{i-1/2,j,k} } {\Delta x^2} +
    \frac{\cF^{3}_{i,j,k+1/2} -\cF^{3}_{i,j,k-1/2} } {\Delta x^3} \, .
\label{DF}
\eeq
As the result Eq.\ref{CONS} becomes a set of coupled ODEs   
\begin{equation}
  \oder{\cQ_{i,j,k} }{t}  = \cL_{i,j,k} (\cQ) 
\label{CONS_a}
\end{equation}
where 

\beq
   \cL_{i,j,k}(\cQ) = -\cDF_{i,j,k}(\cQ) + S(\cQ_{i,j,k}) \, .
\eeq
In this equation, 
the notation $\cDF_{i,j,k}(\cQ)$ means that, strictly speaking, the argument of $\cDF_{i,j,k}$ is 
the whole of the discrete function $\cQ$ (exactly the same applies to $\cL_{i,j,k} (\cQ)$). 
However in reality, only few of its values, in the cells around the $(i,j,k)$-cell, are 
involved -- this defines the scheme stencil.  

It is easy to see that, when the equations of the system (\ref{CONS_a}) are 
integrated simultaneously, the quantities 
\beq
     \sum\limits_{i,j,k} \cQ_{i,j,k}  
\eeq
are conserved down to rounding error. Any larger variation may only be due to 
a non-vanishing source term in (\ref{CONS_a}) or a non-vanishing total flux of the quantity 
through the boundary of the computational domain. In this sense, this finite 
difference scheme is conservative.

\subsection{Third-order WENO interpolation}
    
In order to calculate fluxes at the cell interfaces, and hence $\cDF_{i,j,k}(\cQ)$,
we use the third-order weighted essentially non-oscillatory (WENO) interpolation 
developed by \citet{LOC94}, in the improved form proposed by \citet{YaCa09}. However, 
instead of interpolating the conserved variables we prefer to interpolate the primitive 
ones, $ \cP =(n_\pm,\,p_\pm,\,u^s_\pm,\,E^s,\,B^s,\,\Phi,\,\Psi)$. The benefit of this is 
twofold. Firstly, this reduces by the factor of two the number of required conversions of 
conserved variables into the primitive ones. Indeed, in order to compute the flux vector 
from the vector of conserved variables one would have to convert the latter into the 
vector of primitive variables first. This would have to be done twice per each interface.  
Secondly, this ensures physically meaningful interpolated phase states.    

The aim of the interpolation is to provide us with phase states immediately to the left and 
right of each cell interface given those at the cell centers. In the third order WENO interpolation, 
the interface values are found via a convex linear combination of two linear interpolants, which 
provides third order accuracy on smooth solutions. For the $i$-th cell the interpolants are   
$$
      \cP_{r}(x) = \cP_i+\frac{\Delta \cP_r}{\Delta x}(x-x_i) \etext{and}  
      \cP_{l}(x) = \cP_i+\frac{\Delta \cP_l}{\Delta x}(x-x_i) \, ,  
$$    
where $\Delta \cP_r = \cP_{i+1}-\cP_i$ and $\Delta \cP_l = \cP_{i}-\cP_{i-1}$.  
Here we consider only one of the coordinate directions 
and omit indexes for others. Their convex combination is 
$$
\cP(x) = w_r\cP_r(x) +  w_l\cP_l(x)\, ,
$$
where $w_{r,l}>0$ and $w_r + w_l =1$. These conditions on the weights are satisfied when 
$$
     w_r= \frac{\alpha_r}{\alpha_r + \alpha_l} \etext{and} 
     w_l= \frac{\alpha_l}{\alpha_r + \alpha_l} \,,
$$
where $\alpha_{l,r} >0$.
Removing the arbitrary small constant $\epsilon$, artificially introduced in \citet{YaCa09} in 
order to avoid division by zero during the calculations of $\alpha_{r,l}$, 
and somewhat rearranging their equations, we find the interpolated values of $\cP$ at the 
left and right interfaces of the cell 

\beq
   \cP_{i\pm1/2} = \cP_i \pm(a_{r,l} \Delta \cP_{r,l} + \frac{1}{2}a_{l,r} \Delta \cP_{l,r})/ 
                 (2a_{r,l}+a_{l,r})\, ,
\label{weno_1}
\eeq   
where

\beq
    a_{r,l}= \Delta \cP_{l,r}^2 
\left( \Delta \cP_{r,l}^2 +(\Delta \cP_r + \Delta \cP_l)^2 \right) \, . 
\label{weno_2}
\eeq
It is easy to verify that Eq.\ref{weno_1} yields $\cP_{i\pm1/2} = \cP_i$ 
when either $\Delta \cP_l=0$ or $\Delta \cP_r=0$.  
Thus, one can put 

\beq
   \cP_{i\pm1/2} = \cP_i \etext{if} \Delta \cP_l \Delta \cP_r=0\, . 
\eeq
This allows us to avoid the $0/0$ indeterminacy implied by Eqs.\ref{weno_1},\ref{weno_2}, 
via a simple inspection of data.       

\subsection{Riemann solver}

Using the interpolation described in the previous section, one obtains two phase states, $\cP\sub{L}$ 
and $\cP\sub{R}$, for each cell interface. The state $\cP\sub{L}$ is obtained via the WENO 
interpolation inside the cell to the left of the interface, as described in the previous section, 
where it coresponds to $\cP_{i+1/2}$, whereas the state $\cP\sub{R}$ is obtained via the 
interpolation inside the cell to the right of the interface, where it coresponds to $\cP_{i-1/2}$.       
This defines 1D Riemann problems whose solutions can be used 
to compute the interface fluxes required in Eq.\ref{DF}. Finding exact and even linearized solutions 
of Riemann problems for complicated systems of equations can be rather expensive.  In such cases,  
it makes sense to utilize methods which use minimum amount of information about there characteristic
properties, like the Lax-Friedrich flux, which requires to know only the highest wave speed of the 
system and purely for the stability consideration. Since in the relativistic framework the highest 
speed equals to the speed of light, the latter can replace the former in the Lax-Friedrich 
flux, leading to the equation 

\beq
   \cF\sub{LF} = \frac{1}{2} (\cF\sub{L} + \cF\sub{R}) - \frac{1}{2} (\cQ\sub{R} - \cQ\sub{L})\, ,  
\eeq      
where $\{\cF\sub{L},\cF\sub{R}\}$ and $\{\cQ\sub{L},\cQ\sub{R}\}$ are the vectors of fluxes and 
conserved variables corresponding to the left and the right phase states respectively 
(e.g. $\cF\sub{L}=\cF(\cP\sub{L}) $). 
The second term in the right-hand side of this equation introduces diffusion, which stabilizes 
numerical solutions.  Although this diffusion may be excessive, its negative effect is 
significantly diminished in smooth regions, when schemes with highly accurate reconstruction 
are employed.        

\subsection{Time integration} 

For the integration of the coupled ODEs given by Eq.\ref{CONS_a} we use the popular 
third-order total variation diminishing (TVD) Runge-Kutta scheme by \citet{ShOs88}. 
The one time-step calculations proceed as follows. 

\begin{eqnarray}
\nonumber
  \cQ^{(1)} &=& \cQ^n + \Delta t \cL(\cQ^{n}) \, , \\
  \cQ^{(2)} &=& \cQ^n + \frac{\Delta t}{4} 
               \left(\cL(\cQ^{(1)})+ \cL(\cQ^{n})\right) \, , \\
\nonumber
  \cQ^{n+1} &=& \cQ^n + \frac{\Delta t}{6} 
   \left(\cL(\cQ^{n})+ \cL(\cQ^{(1)}) + 4 \cL(\cQ^{(2)})\right) \, ,
\end{eqnarray}
where $\cQ^n$ and $\cQ^{n+1}$ are the values of $\cQ$ at the times 
$t^n$ and $t^{n+1}=t^n+\Delta t$, and $\cQ^{(1)}$ and $\cQ^{(2)}$ 
are two auxiliary vectors. For stability, the time step must satisfy 
the Courant condition. Since the highest speed of communication in our 
case is the speed of light, this condition reads $\Delta t = c\sub{n} \min(\Delta x^i)$, 
where $c\sub{n} < 1$. Even smaller time step is required when the source terms are 
large.

\subsection{Conversion of conservative into primitive variables} 

The recovering of the hydrodynamic primitive variables is carried out in three steps.  
First we use updated $\bE$ and $\bB$ to remove the electromagnetic contributions 
from the updated total energy and momentum. Second, via adding and subtracting the 
corresponding conserved variables we recover the energies and momenta of the electron 
and positron fluids. As the result, for each of the fluids we have the set of five 
conserved quantities  

\beq
   \cD=n\gamma,\quad \cE = w\gamma^2-p, \quad \cM^i=w\gamma^2v^i\,,
\eeq
where we dropped $\pm$. This is a set of non-linear equation for the primitive 
variables $n$, $p$, and $v^i$. In the literature, one can find a number of different 
algorithms for solving these equations. We employ the following simple iterative 
procedure, developed earlier for our code for relativistic gas dynamics 
\citep{FK96}. Denoting $w\gamma^2$ as $Y$, 
\beq
   Y_{k+1} = \cE+p_k, \quad v^2_{k+1}=\cM^2/Y_{k+1}^2, \quad
   n_{k+1}=\cD/\gamma_{k+1},\quad w_{k+1}=Y_{k+1}/\gamma^2_{k+1},\quad
   p_{k+1}=p(w_{k+1},n_{k+1})\,.  
\eeq  
As the initial guess for $p$ we use its value from the previous time step.
Depending on how good this initial guess is, it takes between 4 iterations in smooth 
regions to 20 iterations near shocks for the solution to converge down to eight 
significant digits.   

\begin{figure*}
\includegraphics[width=80mm]{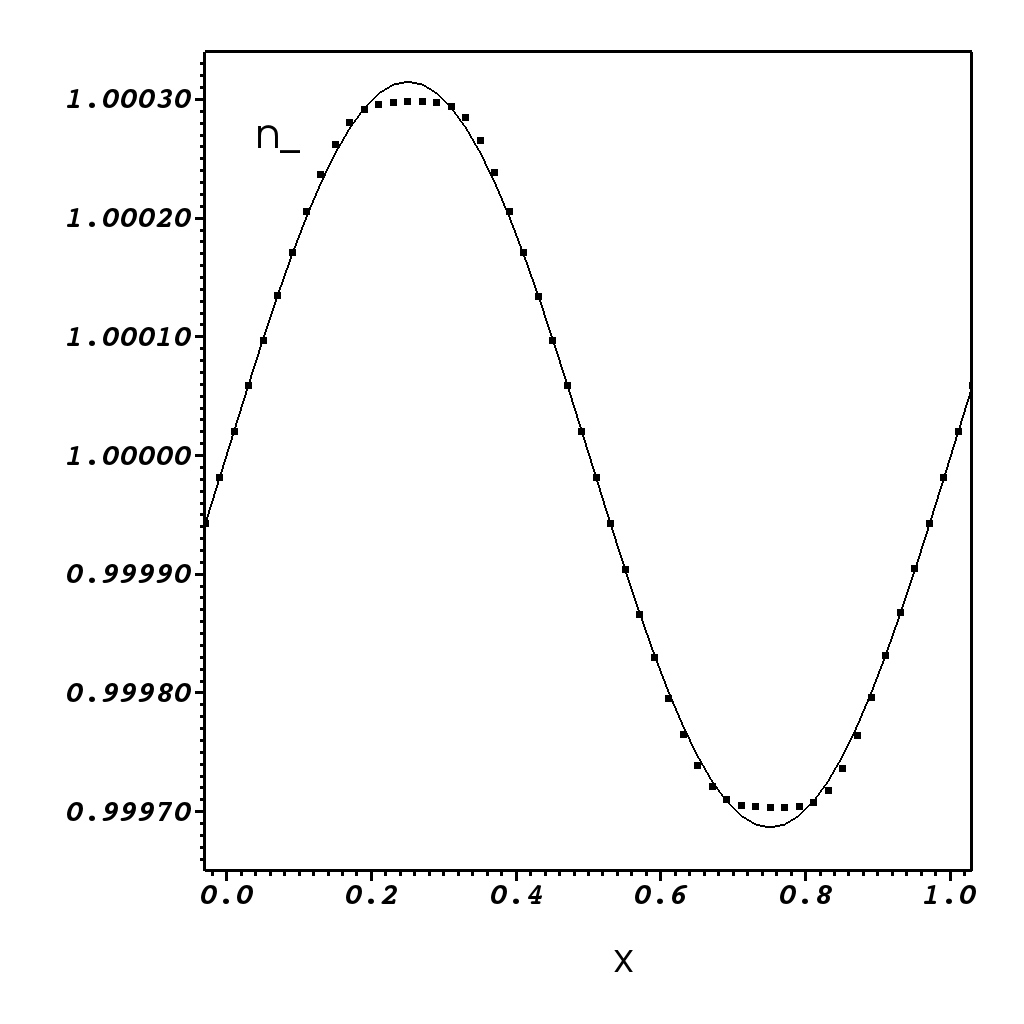}
\includegraphics[width=80mm]{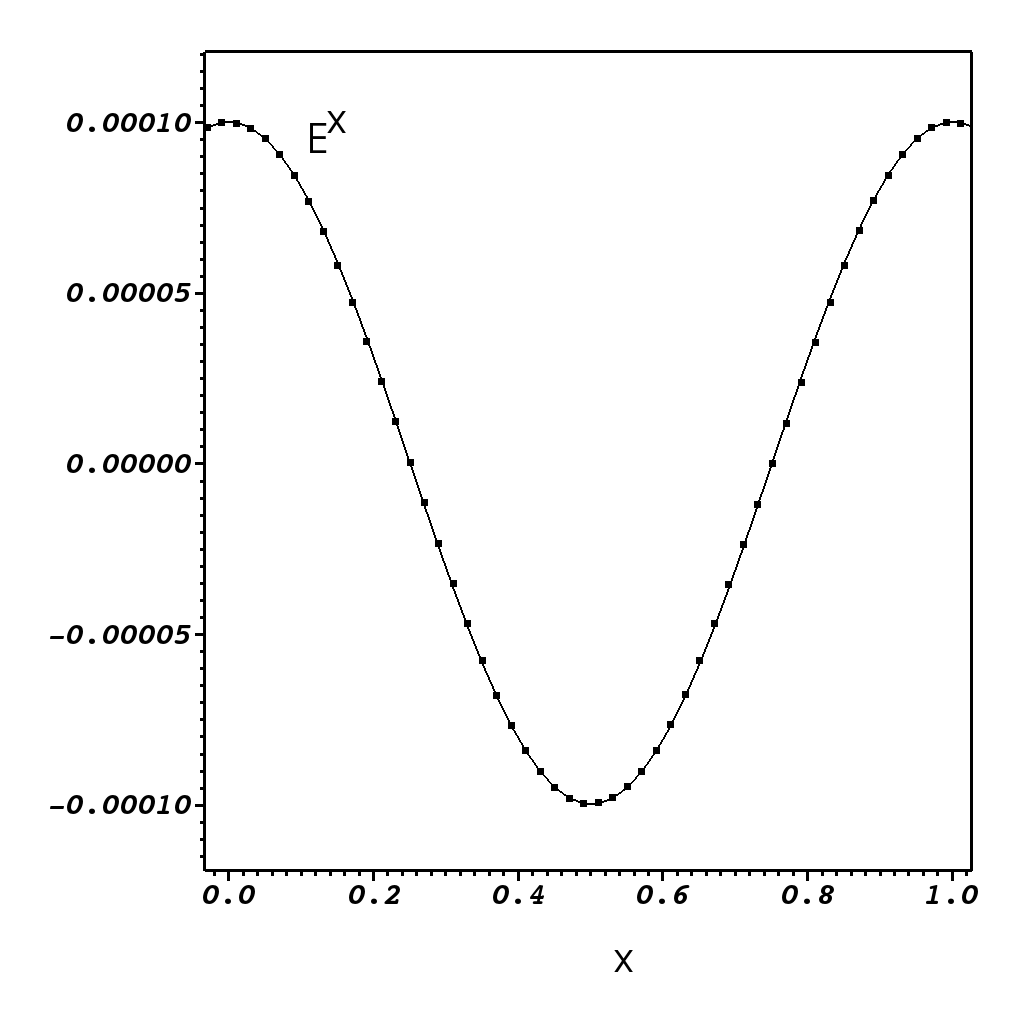}
\caption{ Propagation of small amplitude electrostatic waves. The number density of 
electrons (left panel) and the electric field (right panel). The continuous lines 
show the initial solution and the dots show the numerical solution after one period.  
}
\label{fig:esw}
\end{figure*}

\section{Test simulations} 
\label{tests}

Many of our test simulations are based on slab-symmetric one-dimensional 
problems with known exact analytic or semi-analytic solutions. 
Such problems allow to test not only the 1D version of the code but also 
the multi-dimensional ones. The latter is achieved simply via using the 
multi-dimensional codes to solve one-dimensional problems aligned with one  
of the grid directions. In addition, we used truly multi-dimensional problems for 
which there is no exact solutions but which have been found useful for 
testing one-fluid RMHD codes. The simulations have been carried out either on 
a laptop or/and a workstation with multi-core processors.             

In all simulations, the parameter $\kappa$ of the augmented Maxwell equations 
is fixed at $\kappa = 0.1$. 

\subsection{Electrostatic waves} 

The two-fluid equations allow additional types of waves compared to the single 
fluid MHD. The most basic new waves are the electrostatic waves. In relativistic 
two-fluid MHD with $\Kf=\infty$, 
the properties of linear electrostatic waves were first 
studied by \citet{Sakai80}. Their background solution describes 
uniform state with zero velocity and electric field. The waves propagate along 
the magnetic field and perturb only the particle density and the components of velocity and 
electric field along the magnetic field. Choosing the x axis aligned 
with the magnetic field, putting $A=\tilde{A}\exp(\mbox{i}(\Omega t-kx))$ for the 
perturbation and assuming the same thermodynamic parameters for electron and positron 
fluids, one finds the dispersion relation 

\beq
\Omega^2=\Omega_{p}^2 (1+ (k/k\sub{b})^2) \, , 
\eeq    
where $\Omega_{p}$ is the electron plasma frequency of the uniform background and  
\beq
   k\sub{b}^2= \frac{1}{\Km\Kq} \fracp{2 n^2\sub{b}}{\Gamma p\sub{b}} \, ,
\eeq
where the $n\sub{b}$ and $p\sub{b}$ are the background electron (and positron) number 
density and pressure respectively. 
For the Fourier amplitudes one has 
\beq
\tilde{n}_\pm = n\sub{b} \fracp{k}{\Omega} \tilde{v}^x_\pm \, 
\eeq   
and 

\beq
    \tilde{v}^x_{\pm} = \pm \mbox{i} \frac{1}{\Km} 
    \frac{n\sub{b}\Omega}{(\Gamma p\sub{b}k^2 - w\sub{b}\Omega^2)} \tilde{E}^x \, .
\label{v_amp}
\eeq 
Figure \ref{fig:esw} shows the results of 1D test simulations based on this 
solution. The parameters used in this test are $\Kq=\Km=1$, $\Kf=\infty$, $k=2\pi$, 
$\Gamma=4/3$, $n\sub{b}=p\sub{b}=1$ and $\tilde{E}^x=10^{-4}$. 
The uniform computational grid has 50 cells covering the domain $[0,1]$. 
One can see that in spite of the relatively 
low spatial resolution, the numerical solution is quite accurate, particularly for the 
electromagnetic field. Noticeable errors are observed only in the gas parameters 
near the turning points of the wave curve.

\begin{figure*}
\includegraphics[width=80mm]{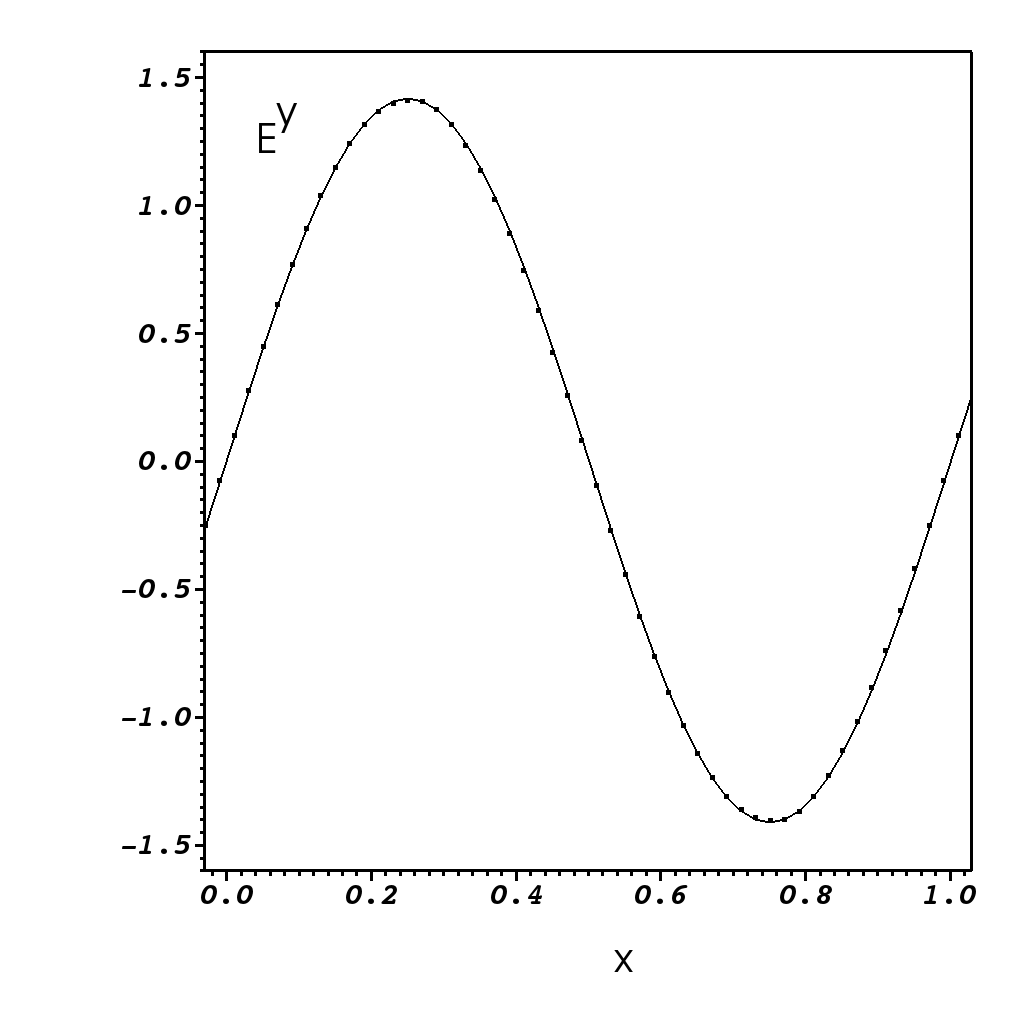}
\includegraphics[width=80mm]{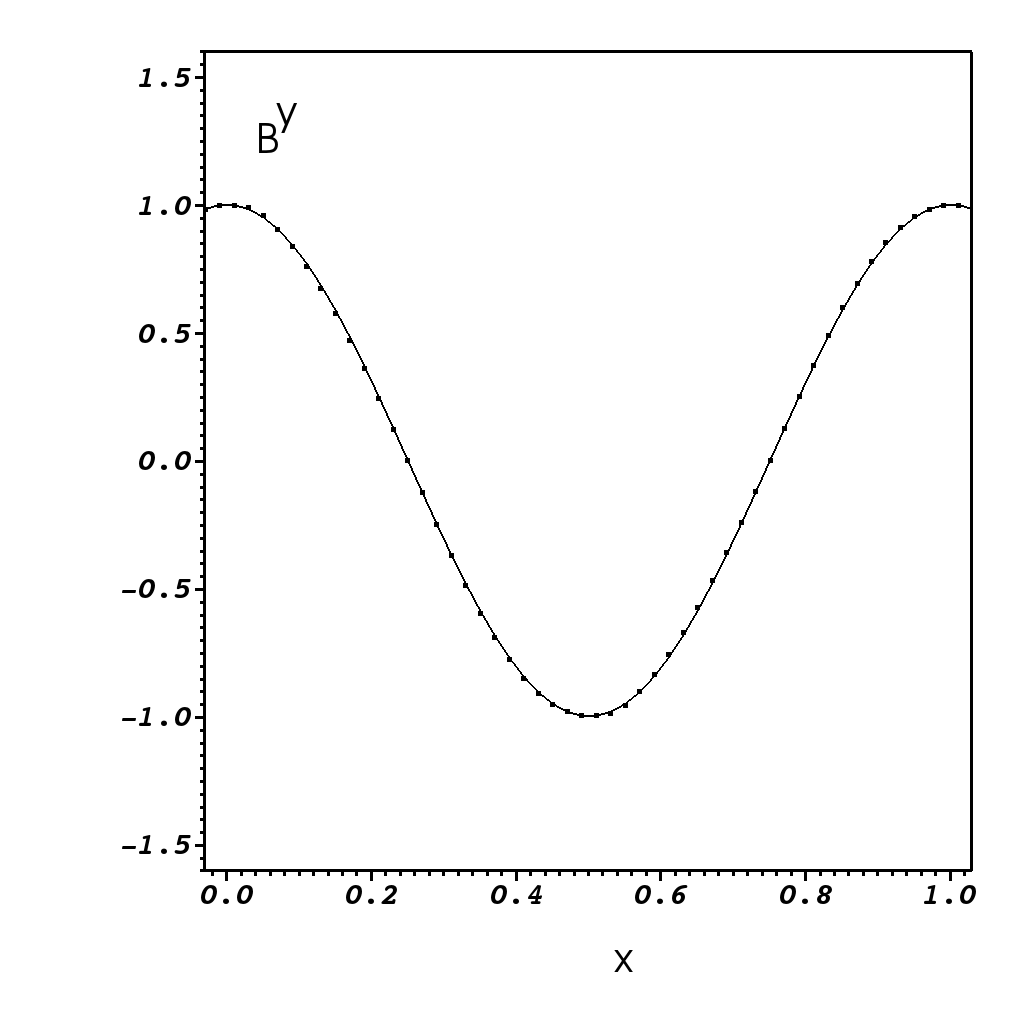}
\caption{ Propagation of circularly polarized ``superluminal''  waves. The left panel 
shows $E^y$ and the right panel shows $B^y$. The continuous lines
show the initial solution and the dots show the numerical solution after one period.
}
\label{fig:cpw}
\end{figure*}

\begin{table}
\begin{center}
\caption{L1 norm errors in the problem of circularly polarized wave.}
    \label{cpw-l1}
   \begin{tabular}{|c|c|c|c|}
   \hline
   \hline
       $N$ & $p_{-}$ & $B$ & $u_{-}$ \\    
   \hline
   \hline
     50 & $9.94\times10^{-2}$  &  $2.73\times10^{-3}$ & $3.56\times10^{-1}$ \\
   \hline
     100 & $1.07\times10^{-2}$  &  $4.23\times10^{-4}$ & $4.55\times10^{-2}$ \\
   \hline
     200 & $1.23\times10^{-3}$  &  $5.81\times10^{-5}$ & $5.32\times10^{-3}$ \\
   \hline
     400 & $1.46\times10^{-4}$  &  $7.57\times10^{-6}$ & $6.34\times10^{-4}$ \\
   \hline
   \hline
   \end{tabular}
\end{center}
\end{table}

\subsection{Circularly polarized ``superluminal'' waves.} 

These are nonlinear harmonic solutions of the two-fluid equations with 
$\Kf=\infty$ \citep[e.g.][]{Max73,Amano13}. Their background solution describes
uniform state with vanishing velocity, magnetic and electric fields and equal densities 
and pressures of the electron and positron fluids. 
These waves do not perturb particle density and pressure, as well as the velocity 
component along the wave vector. The components of vectors normal to the wave 
vector display pure rotation, whose direction depends on the wave polarization.      
Choosing the x-axis aligned with the wave vector 
and putting $A=\tilde{A}\exp(\mbox{i}(\Omega t -kx))$ for the wave of left 
polarization one finds the dispersion relation 

\beq
\Omega=\Omega\sub{p}(1+(k/k\sub{b})^2)^{1/2},
\eeq
where $\Omega_p$ is the plasma frequency and 
$k\sub{b}=1/\Omega\sub{p}$. The corresponding Fourier amplitudes are related via 
\beq
    \tilde{\bu}_{\perp,\pm} = \pm \frac{1}{\Km}
    \frac{n\sub{b}}{kw\sub{b}} \tilde{\bB}_{\perp} \, , 
\label{v_cpw_amp}
\eeq

\beq
    \tilde{\bE}_{\perp} = - \mbox{i} \frac{\Omega}{k} \tilde{\bB}_{\perp} \, ,  
\label{e_cpw_amp}
\eeq
where as before $n\sub{b}$ and $w\sub{b}$ are the unperturbed background electron density 
and enthalpy respectively. 
For the right polarization, $A=\tilde{A}\exp(-\mbox{i}(kx-\Omega t))$, 
Eq.\ref{v_cpw_amp} is replaced with 
\beq
    \tilde{\bu}_{\perp,\pm} = \mp \frac{1}{\Km}
    \frac{n\sub{b}}{kw\sub{b}} \tilde{\bB}_{\perp} \, .
\label{v_cpw_amp1}
\eeq

Figure \ref{fig:cpw} shows the results of 1D test simulations, based on the 
solution for the left polarization wave. 
The parameters of the problem are $\Kq=1$, $\Km=1/(2\pi)^2$, $\Kf=\infty$, 
$p\sub{b}=1/4$, $n\sub{b}=1$, $\Gamma=4/3$, $v_\pm^x =0$, $B^x=0$ and $k=2\pi$. 
The corresponding wave phase speed is $v_\phi = \sqrt{2}$.
The uniform computational grid has 50 cells and covers the domain $[0,1]$. 
In this figure the continuous lines show the initial solution and the dots show the 
numerical solution after one wave period, at $t=1/\sqrt{2}$. Table \ref{cpw-l1} show the 
L1 norm of the numerical errors for $p_{-}$  and the magnitudes of $B$, and $u_{-}$, all of which 
should remain constant. The data confirm that the scheme is third order accurate.

\begin{figure*}
\includegraphics[width=80mm]{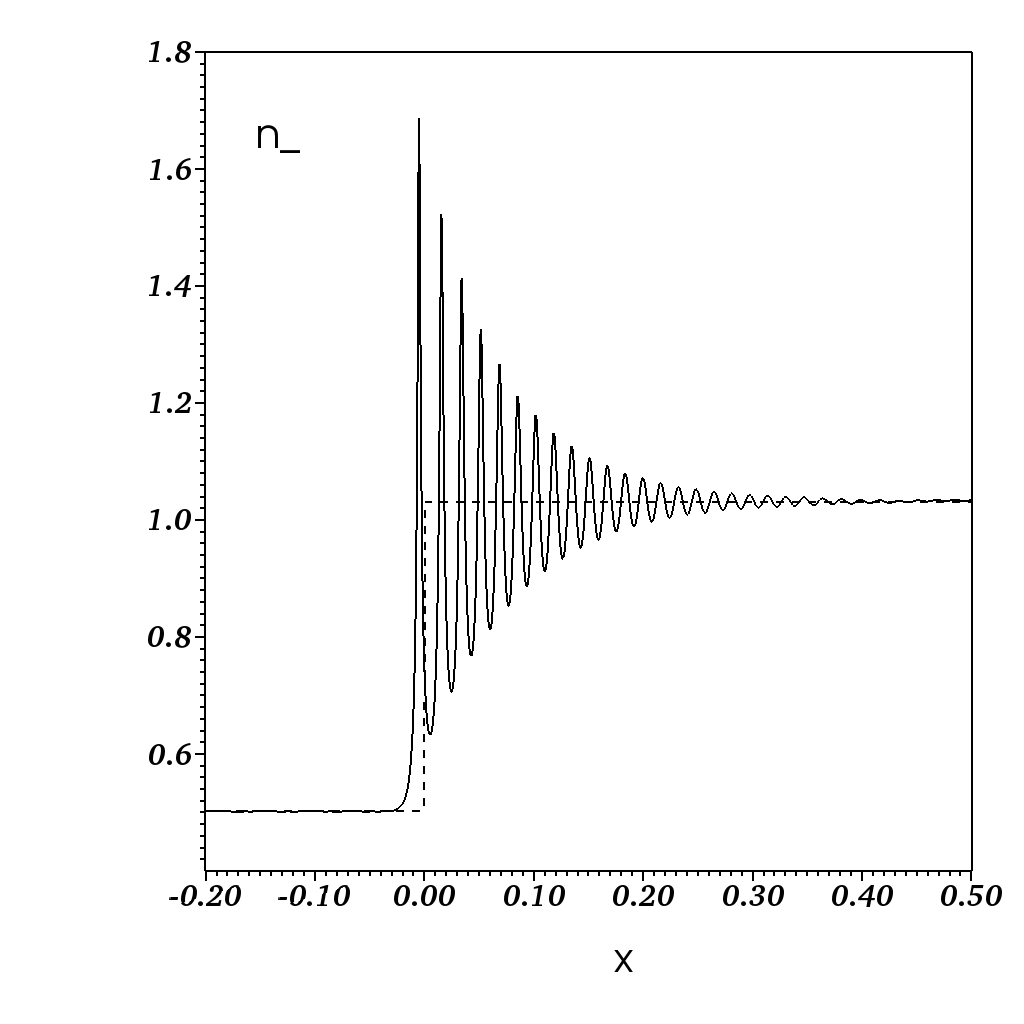}
\includegraphics[width=80mm]{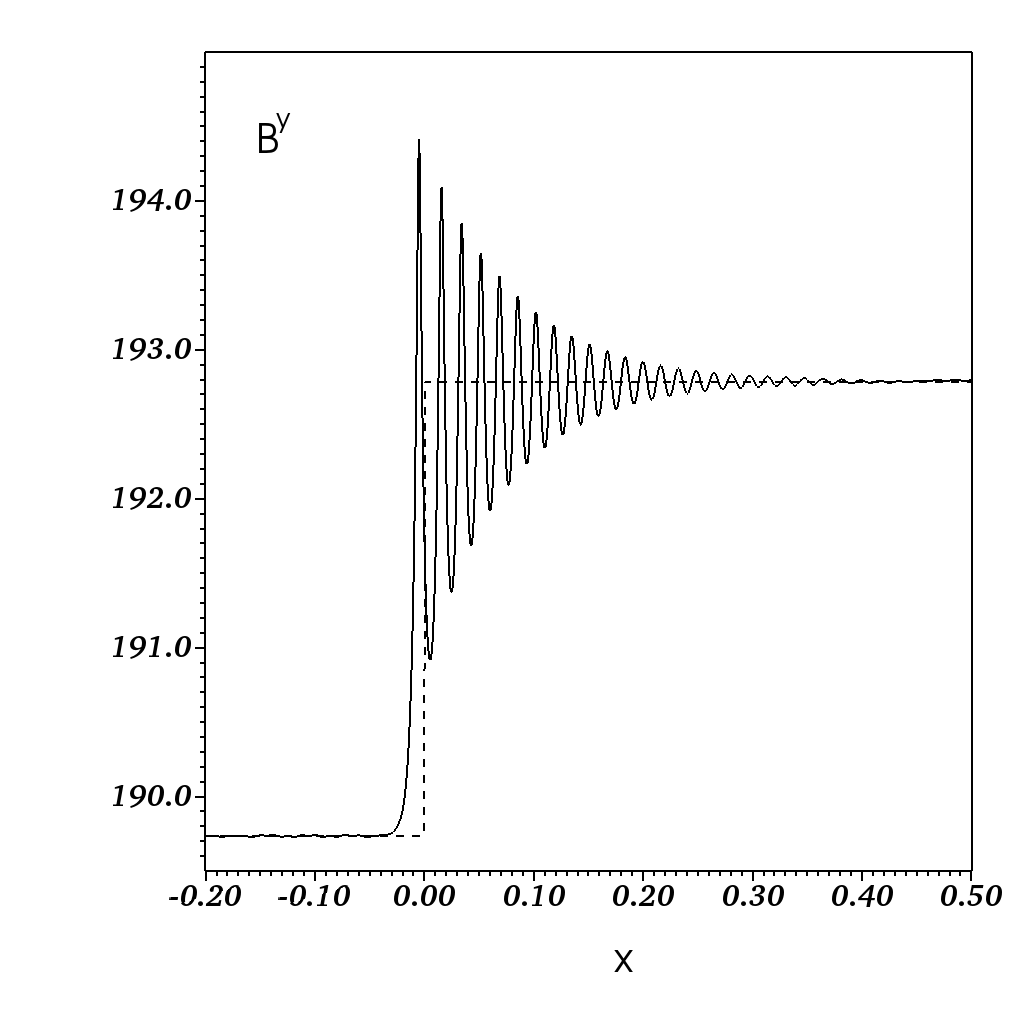}
\includegraphics[width=80mm]{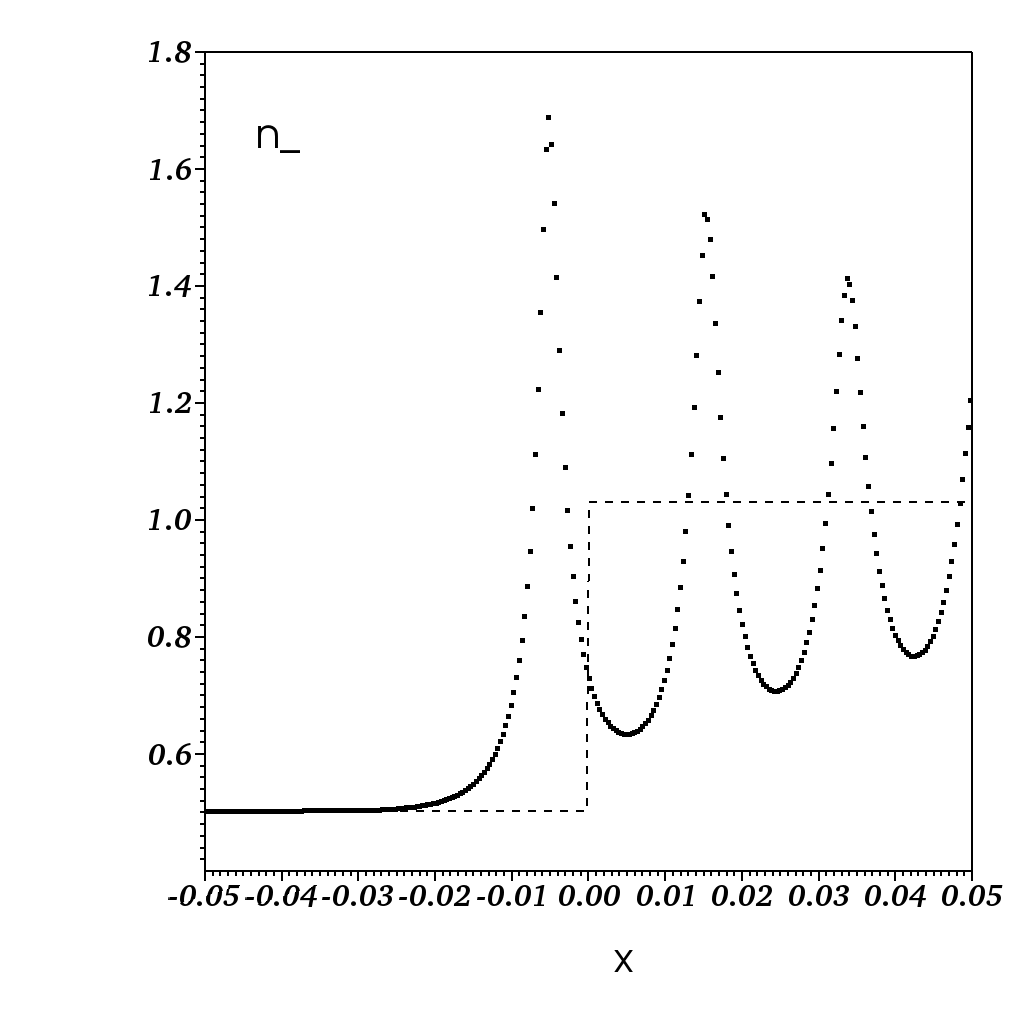}
\includegraphics[width=80mm]{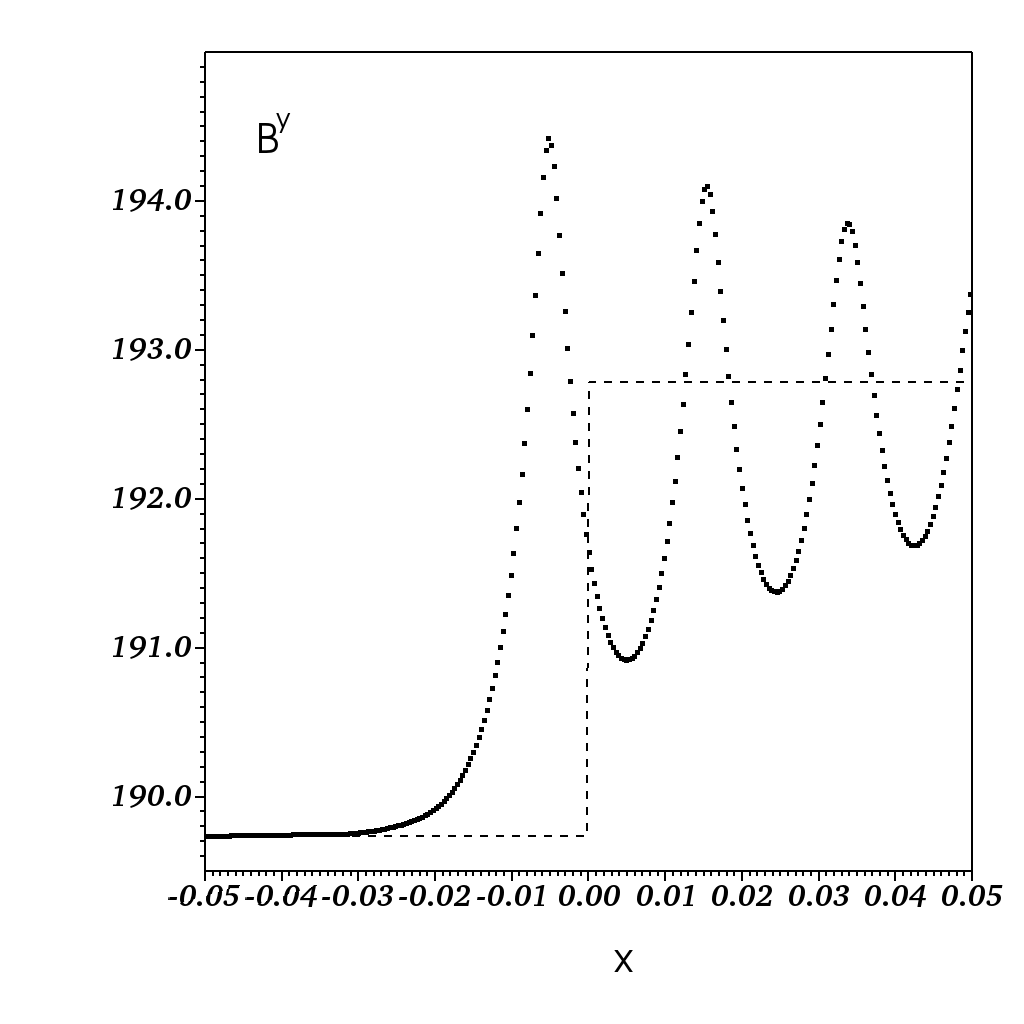}
\caption{ 
Fast magnetosonic shock case 1. The dashed line shows one-fluid RMHD 
shock solution. The solid lines in the top panels and the dots in the 
bottom panels show the numerical two-fluid solution.  
}
\label{fig:fs1}
\end{figure*}

\begin{figure*}
\includegraphics[width=80mm]{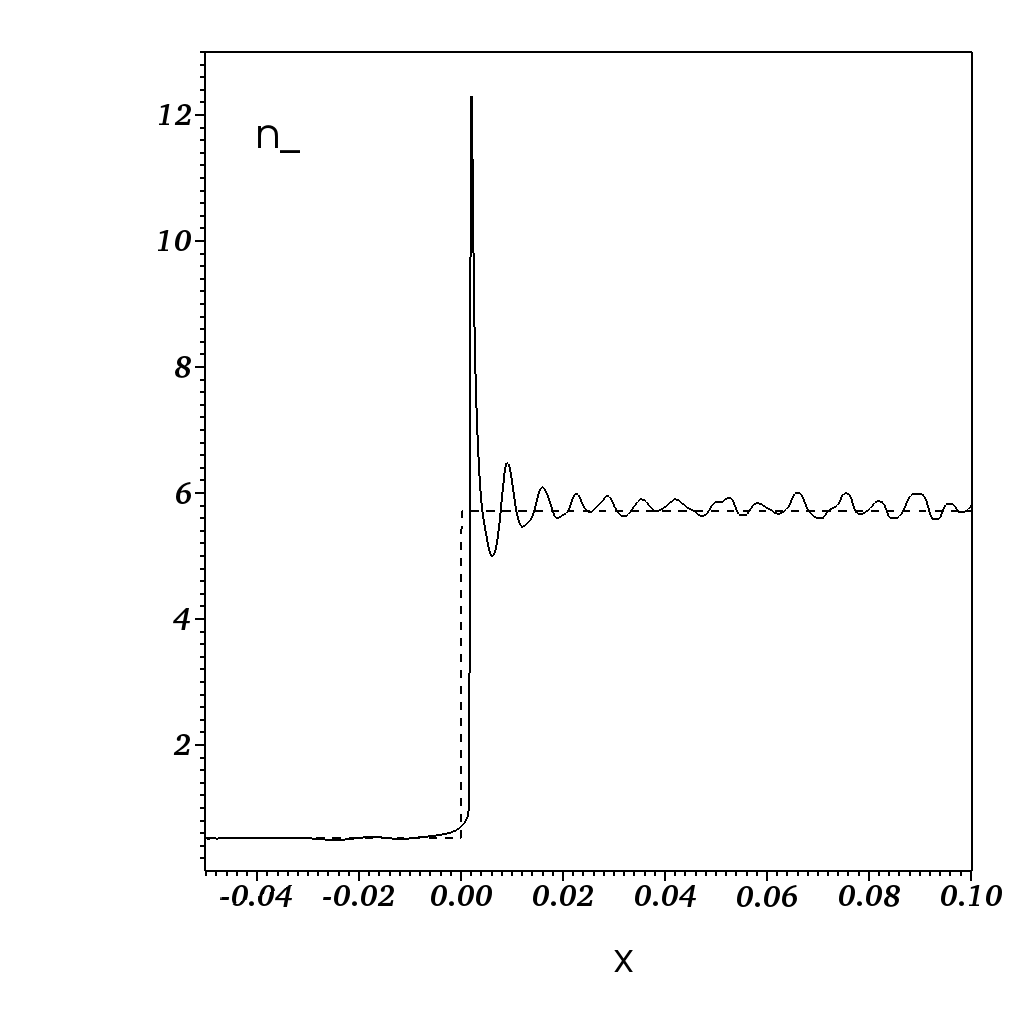}
\includegraphics[width=80mm]{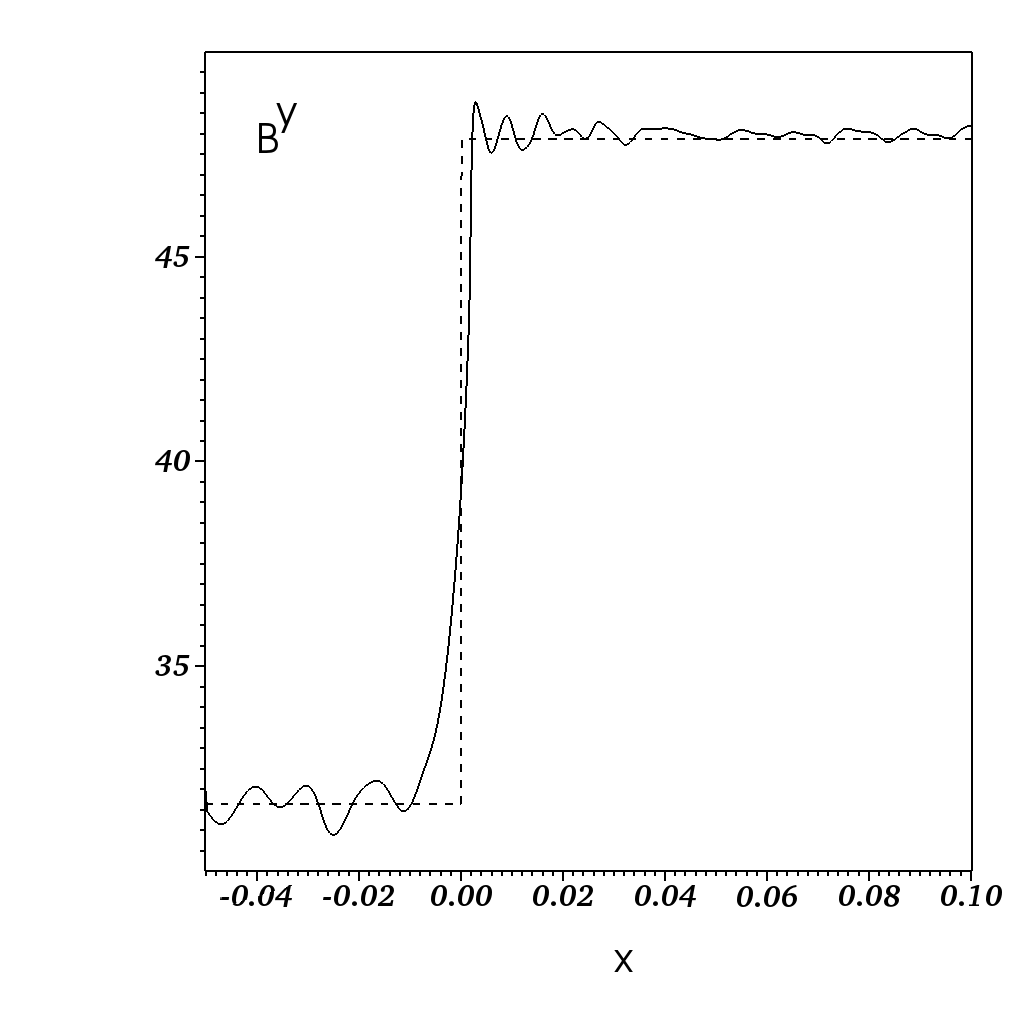}
\includegraphics[width=80mm]{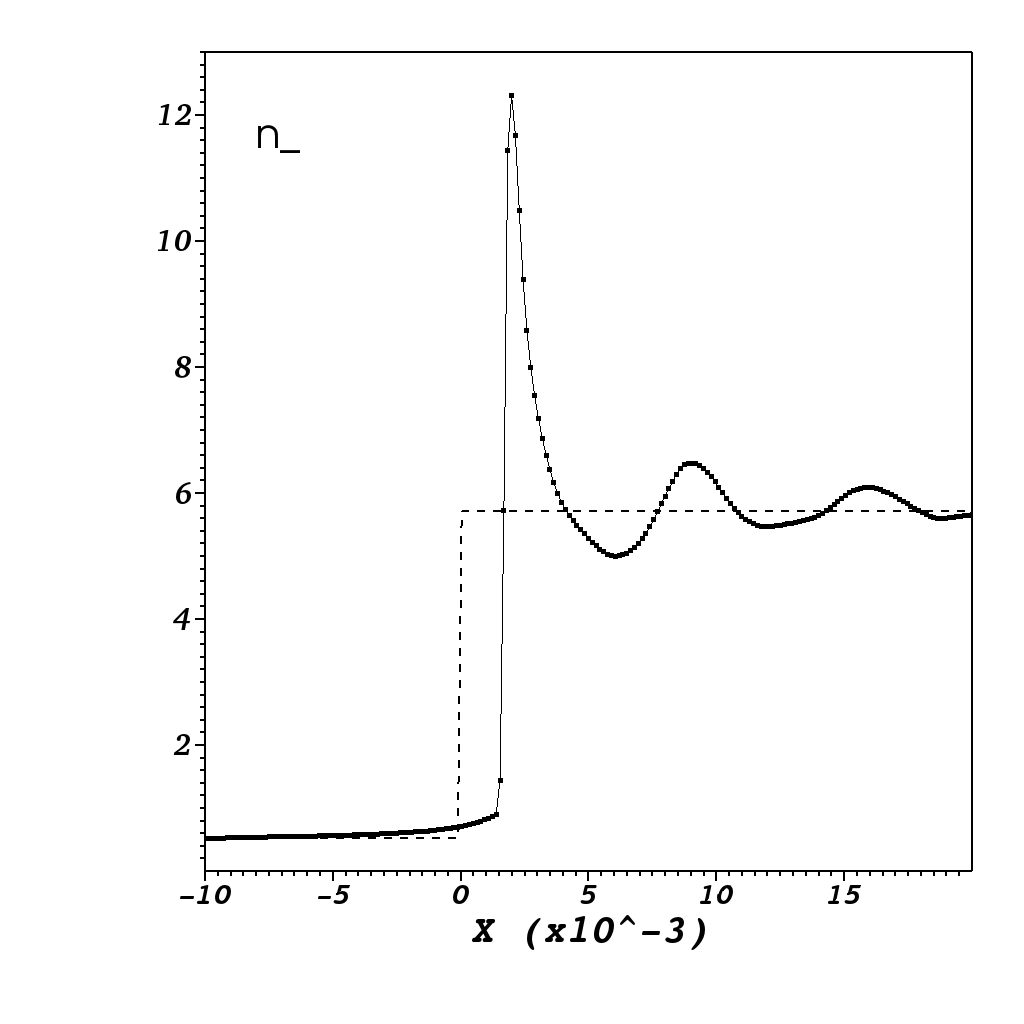}
\includegraphics[width=80mm]{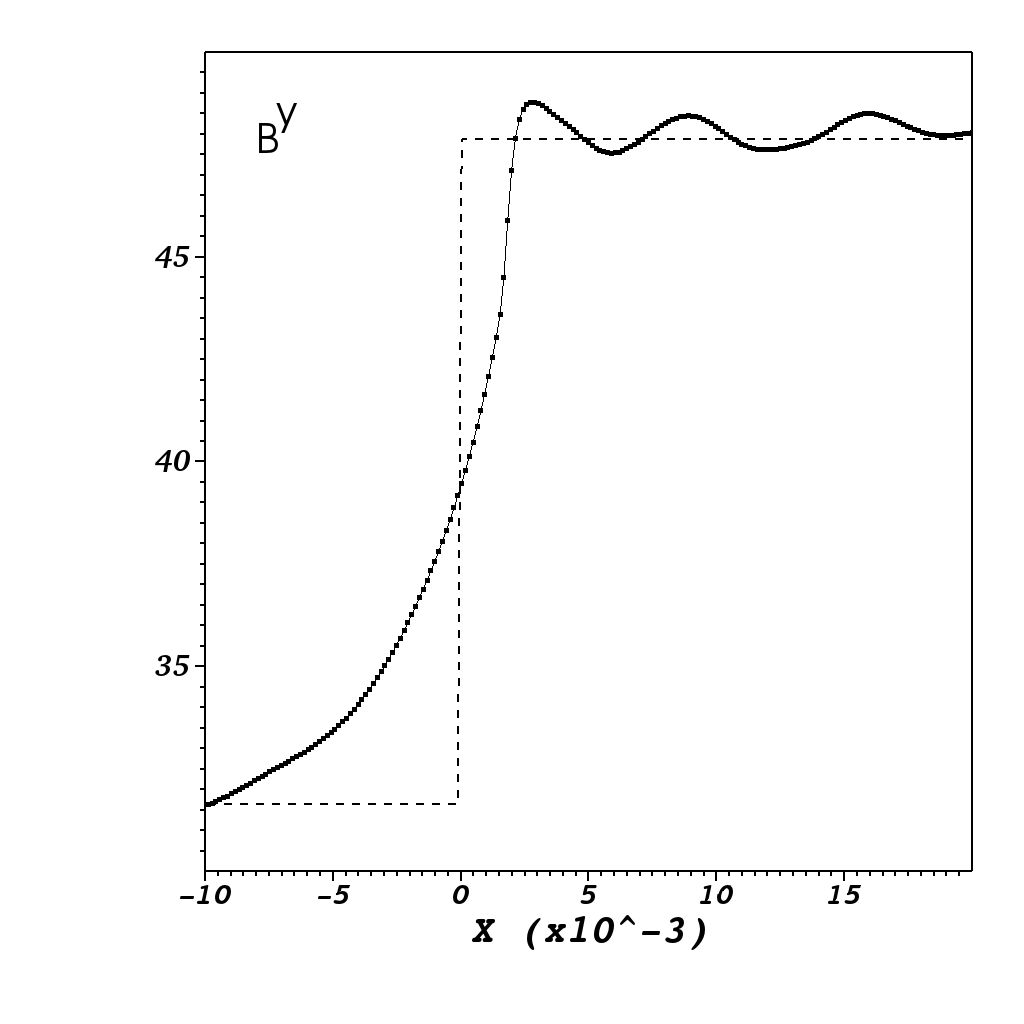}
\caption{
Fast magnetosonic shock case 2. The dashed line shows one-fluid RMHD
shock solution. The solid lines in the top panels and the dots in the
bottom panels show the numerical two-fluid solution.
}
\label{fig:fs2}
\end{figure*}

\begin{table}
\begin{center}
\caption{Perpendicular fast magnetosonic shock solutions used in test 
simulations. The flow Lorentz factor and magnetic field strength are 
given in the shock frame. }\label{fs-tests}
   \begin{tabular}{|c|l|l|}
   \hline
   \hline
       Case  & Left state & Right state \\
   \hline
   \hline
      1 &  $n_-,n_+=0.5$          & $n_-,n_+=1.02977$ \\
        &  $p_-,p_+=0.05$         &  $p_-,p_+=0.171023$ \\
        &  $\gamma_-,\gamma_+=10$    & $\gamma_-,\gamma_+=4.93331$ \\
        &  $B=189.728$    & $B=192.779$ \\
   \hline
      2 &  $n_-,n_+=0.5$          & $n_-,n_+=5.70015$ \\
        &  $p_-,p_+=0.05$         & $p_-,p_+=8.09570$ \\
        &  $\gamma_-,\gamma_+=10$    & $\gamma_-,\gamma_+=1.32728$ \\
        &  $B=31.6214$    & $B=47.8497$ \\
   \hline
   \hline
   \end{tabular}
\end{center}
\end{table}

\subsection{Perpendicular fast magnetosonic shocks.}
\label{pfms}

In these tests, we utilized  single fluid ideal RMHD solutions for 
fast magnetosonic shock waves which were obtained by solving the relevant
shock equations. Although these solutions are discontinuities the corresponding 
two-fluid solutions could have continuous structure with or without sub-shocks. 
In any case, the asymptotic left and right states must be the same as in the 
single fluid solutions and this explains the virtue of such tests.     
The sub-shocks may appear as the two-fluid  equations do not have terms 
corresponding to viscosity. 

Here we present only the case of perpendicular shock, where 
the magnetic field is tangent to and the flow velocity
is normal to the shock front respectively. 
The parameters of the one-fluid solutions in the shock frame 
are given in table~\ref{fs-tests}. They were used to setup discontinuities in 
the initial data ( Riemann problems ). 
In both cases, $\Km=\Kf=0.01$ and $\Kq=0.001$. In the case 1, the numerical 
grid contained  2000 cells covering the domain $[-0.2,0.5]$, whereas in the case 2, 
it had 1000 cells and the domain $[-0.05,0.1]$. As to the boundary 
conditions, we first used the zero gradient ones ( also known as the ``free flow'' 
conditions) at both the boundaries, just like in the similar test simulations for 
our one-fluid RMHD code. However, various waves were emitted from the position of the 
initial discontinuity in both directions\footnote{Waves can propagate upstream of 
fast shocks because in the two-fluid MHD their speed is not limited from above by 
the fast magnetosonic speed.},
noticeably upsetting both the boundary states. This resulted in slightly 
different solutions compared to the reference ones. A somewhat better outcome 
is obtained when the upstream boundary state is fixed at the reference state. 
In any case, both boundaries reflect incident waves and the reflected waves interact 
with the shock, so a certain level of noise is present in the numerical solutions.

\begin{figure*}
\includegraphics[width=80mm]{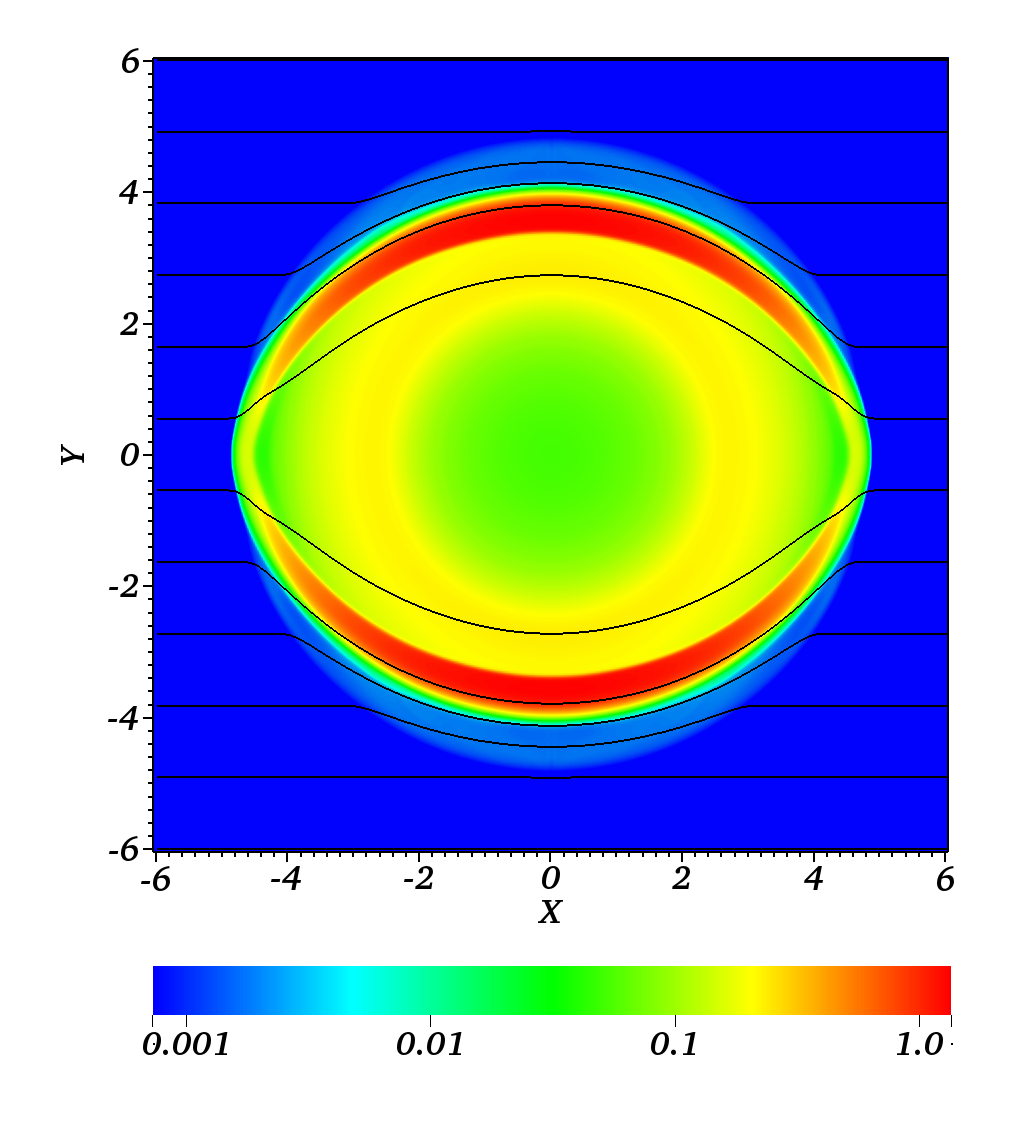}
\includegraphics[width=80mm]{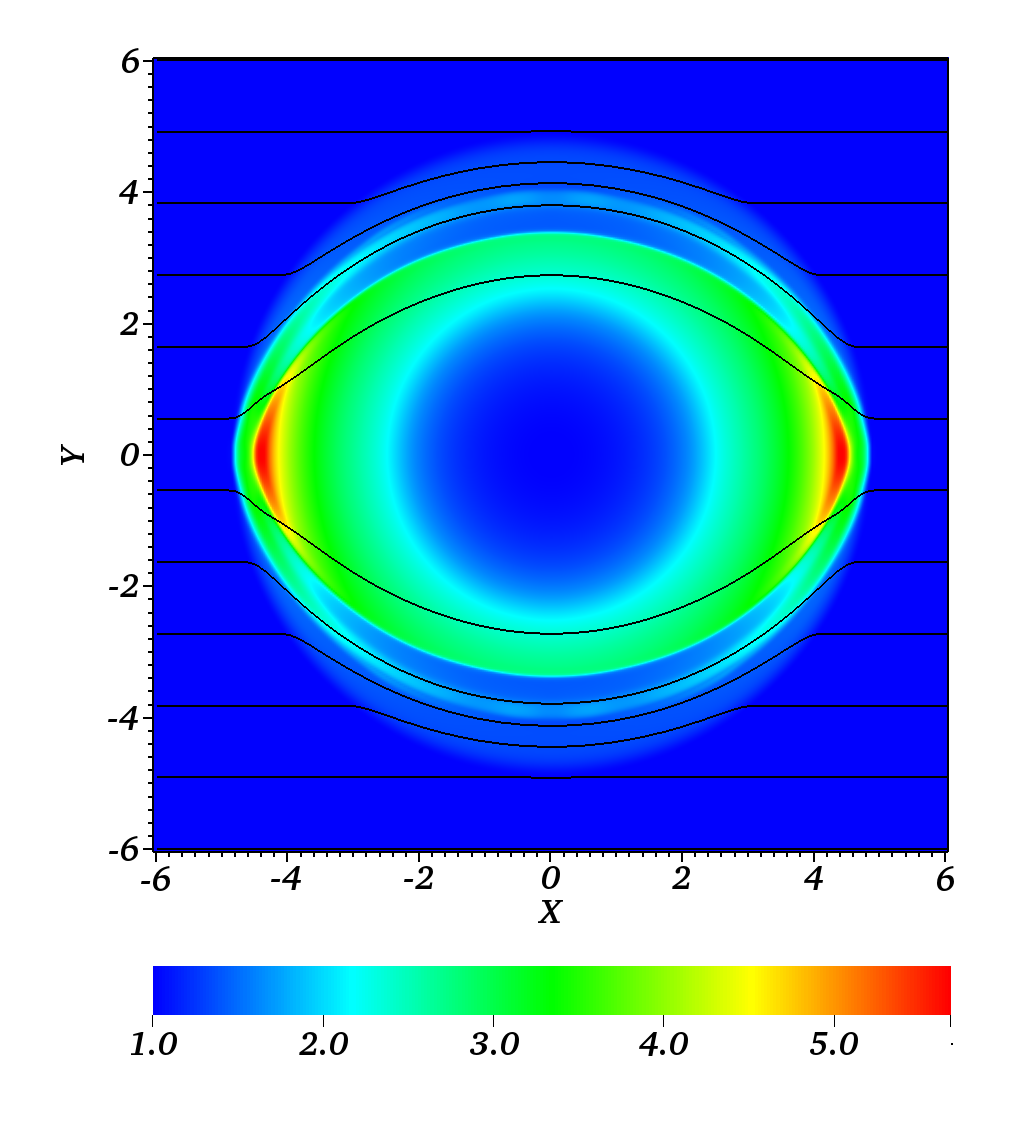}
\includegraphics[width=80mm]{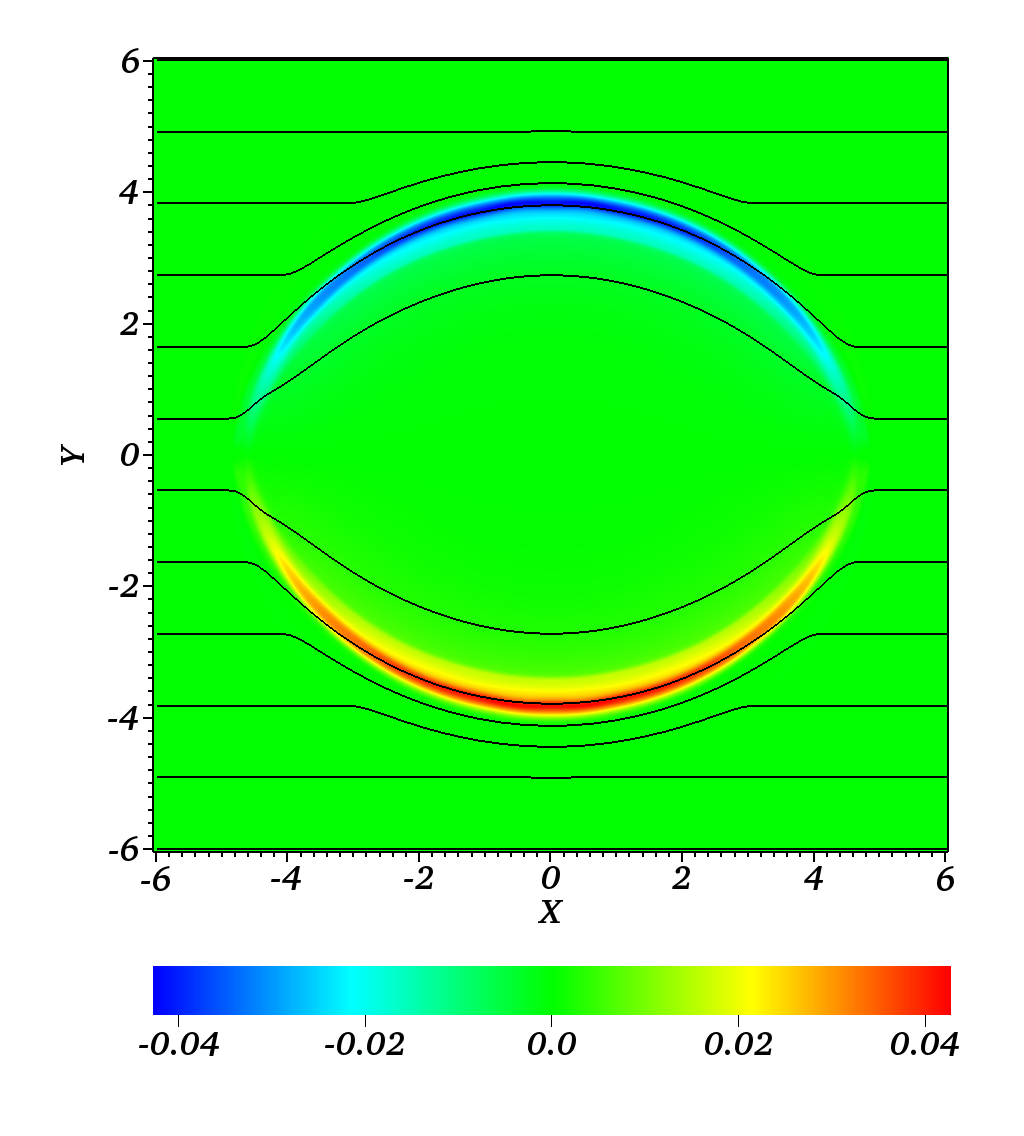}
\includegraphics[width=80mm]{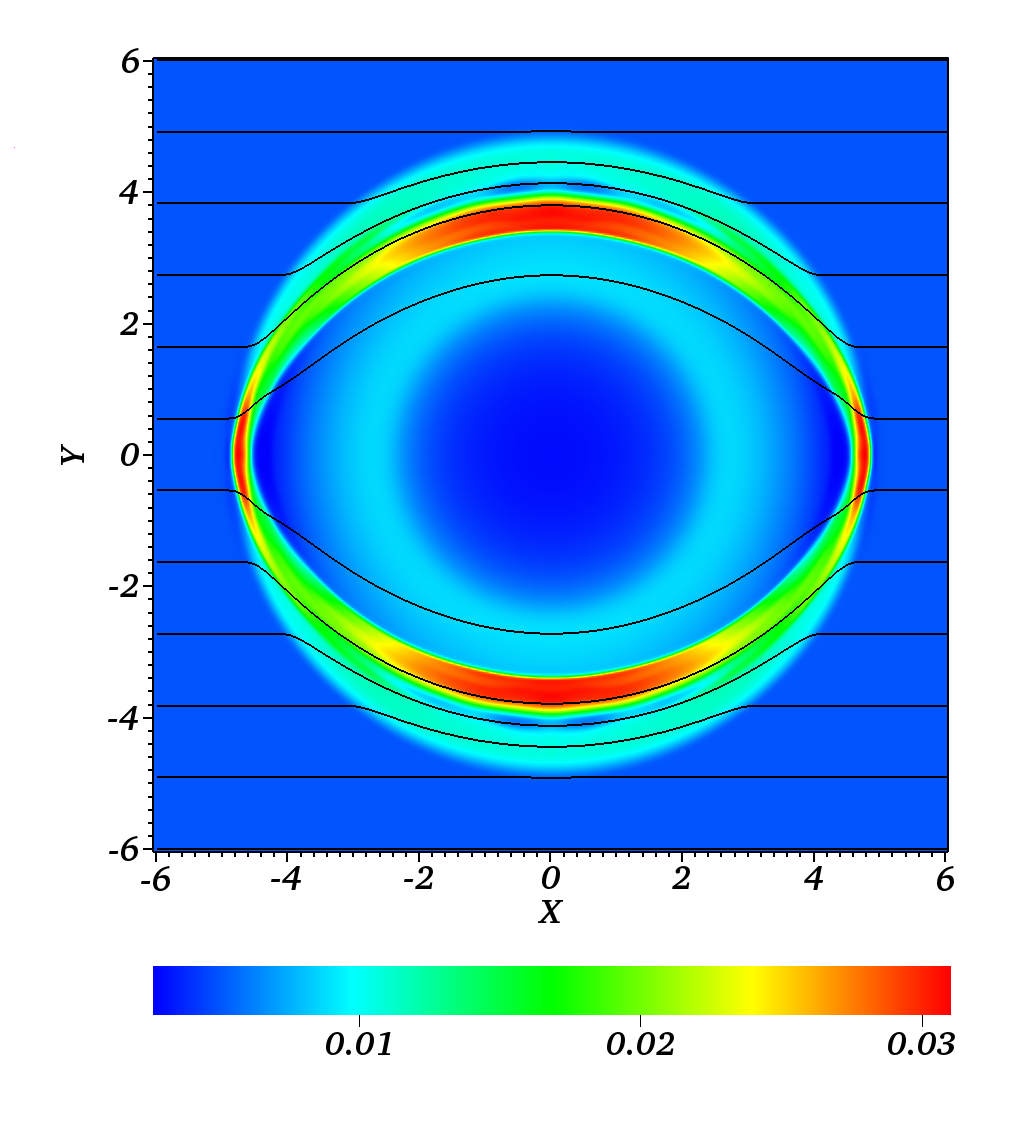}
\caption{
Cylindrical Blast Wave 2D test. Time $t=4$.
The top left panel shows the gas pressure, $\log_{10} p_+=\log_{10} p_-$,
The top right panel shows the Lorentz factor $\gamma_-=\gamma_+$, the maximum value being 
$\gamma\sub{max}=5.685$. 
The bottom left panel shows the z component of electric current, $j^{z}=n_+u^z_+ - n_-u^z_-$. 
The bottom right panel shows the gas density $n_-=n_+$. 
All plots show the magnetic field lines.
}
\label{fig:bw2d}
\end{figure*}

\begin{figure*}
\includegraphics[width=80mm]{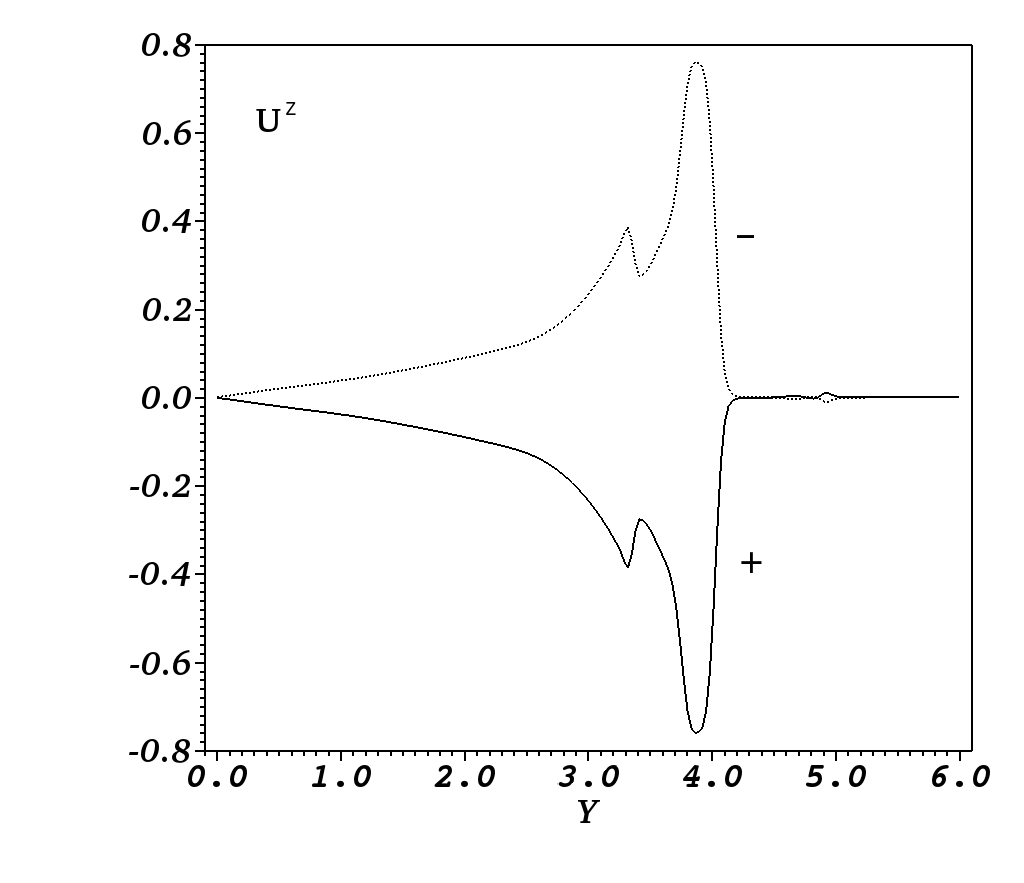}
\caption{
Cylindrical Blast Wave 2D test. Time $t=4$. The solid line shows $u^z_+$ 
and the dotted line $u^z_-$ along the line $x=0$. 
}
\label{fig:bw2d-uz}
\end{figure*}

Figure~\ref{fig:fs1} shows that in the case 1 the 
two-fluid solution does not involve a sub-shock -- it 
is continuous in all variables.  The downstream part of the solution is 
a wave-train of gradually diminishing amplitude. The rate of decay is determined
by the dimensionless parameter $\Kf$, higher $\Kf$ yielding slower decay. 
This suggests that for finite $\Kf$, the fixed point of the shock structure 
equations corresponding to the right state is a stable focus.  
For $\Kf=\infty$, the fixed point must be encircled by a limit cycle. 
This structure is in general agreement with the analysis of nonrelativistic shocks 
by \citet{Sag66}. However, the wavelength of the wavetrain is significantly 
smaller than the skin-depth, which is $\lambda\sub{s}\simeq 3$, probably due to the 
relativistic effects.  On the contrary, the generalized gyration radius, 
$R_g\simeq 7\times10^{-4}$, is small compared to the wavelength\footnote{The equation for
the generalized gyration radius, which is the gyration radius corresponding to the typical 
thermal speed of plasma particles, is given in Table~\ref{pp}.}. 
When the wavelength is not properly resolved, the numerical solution develops strong 
oscillations on the cell scale, leading to a crash, unless these oscillations 
are dumped by high resistivity.   
  
In contrast, the case 2 solution involves a hydro sub-shock --  the 
density, pressure and velocity of the fluids are discontinuous, whereas the 
electric and magnetic fields are not (see figure~\ref{fig:fs2}). 
Downstream of the sub-shock, one can also see a wave-train similar to that of
the case 1 but of smaller amplitude. 
In both cases, the two-fluid solution agrees with the single fluid solution at 
$x\to\pm\infty$.   
In this problem, the generalized gyration radius $R_g\simeq 4\times10^{-3}$ and the 
skin depth is $\lambda\sub{s}\simeq 3$. 

In our test problems, the flow magnetization $\sigma\sub{m} \simeq 13$ and 
$\sigma\sub{m}\simeq 0.36$ for
the cases 1 and 2 respectively. Thus, our results are is in agreement with 
those  of \citet{hgskb12}, who analyzed the structure equations of one-dimensional 
nonlinear perpendicular waves and concluded that continuous soliton-type solutions of 
two-fluid RMHD do not exist in the low $\sigma\sub{m}$ limit.

\subsection{Cylindrical and Spherical Blast Waves}

All the test problems described so far are one-dimensional with 
the slab symmetry. We used them to test not only our 1D scheme but also 
the 2D and 3D schemes. In these multi-dimensional tests, the initial solutions 
were aligned with the grid, so that they varied only along one of the coordinate 
directions. The disadvantage of such approach is that it excludes waves propagating  
at an angle to the grid. To verify that the code handles such cases as well,  
we carried out the following two tests.  

The first one is the so-called cylindrical explosion problem, as introduced in 
\citet{ssk-godun99} for testing ideal RMHD codes and later applied by many other
researches.  This is a 2D problem with the slab symmetry 
and it describes the evolution of a cylindrical fireball in a uniform 
perpendicular magnetic field. The domain is $[-6,6]\times[-6,6]$ with a 
uniform grid of $400\times400$ cells. 
The magnetic field is $B=(1,0,0)$  throughout the domain. The fireball is 
centered on the origin and its initial radius is $r\sub{b}=1$. In the 
surrounding, for $r>r\sub{b}$, the plasma parameters 
are $2n_+=2n_-=0.01$ and $2p_+=2p_-=0.003$. Initially, inside the fireball 
$2n_+=2n_-=1$ and $2p_+=2p_-=100$ for $r<0.8\,r\sub{b}$, and these decay 
exponentially down to the external plasma values for $r\sub{b}<r<0.8\,r\sub{b}$    
The ratio of specific heats $\Gamma=4/3$. Initially, the velocity vanishes 
everywhere. $\Km=\Kq=0.01$ and $\Kf=1$.  The dimensionless generalized gyration radius is 
$R\sub{g} = 0.022$ and $4.0$ outside and inside of the fireball respectively and
the corresponding magnetization $\sigma\sub{m} = 23.$ and $1.2\times10^{-3}$. 

Figure~\ref{fig:bw2d} illustrates the solution obtained with the 
2D code at $t=4$. One can see that angle between the blast wave and the grid 
varies continuously, covering the whole available range, and so does the angle with 
the magnetic field. The latter is important as the properties of waves moving 
along, perpendicular and at an intermediate angle to the magnetic field vary 
significantly. In ideal MHD, well known degeneracies occur along and perpendicular 
to the magnetic field. Overall, 
the two fluid-solution is very similar to the corresponding single fluid ideal
\citep{ssk-godun99} and  resistive \citep{ssk-rrmhd} RMHD solutions, 
with all the key features present and having similar parameters. 
The highest Lorentz factor on the grid, $\gamma\sub{max}=5.685$, is found on the 
$y=0$ line, very close to the reverse shock.  This value is somewhat higher than 
in the single-fluid solutions. Given the very small size of this high speed 
region, this  difference is most likely to be due to the higher accuracy of 
the two-fluid simulations (The single fluid codes were only of the second order 
accuracy.). What is qualitatively different from the single-fluid 
data is the unequal velocities of the electron and positron fluids. 
This is illustrated in Figure~\ref{fig:bw2d-uz} which shows the z component of 
the four-velocity along the y axis. It is this difference in velocities 
which yields the electric current shown in bottom left panel of Figure~\ref{fig:bw2d}. 

This 2D problem was also used to test the 3D code, namely we 
carried out three test runs with the symmetry axis of the fireball aligned 
with one of the three basis directions of the 3D Cartesian grid.

\begin{figure*}
\includegraphics[width=80mm]{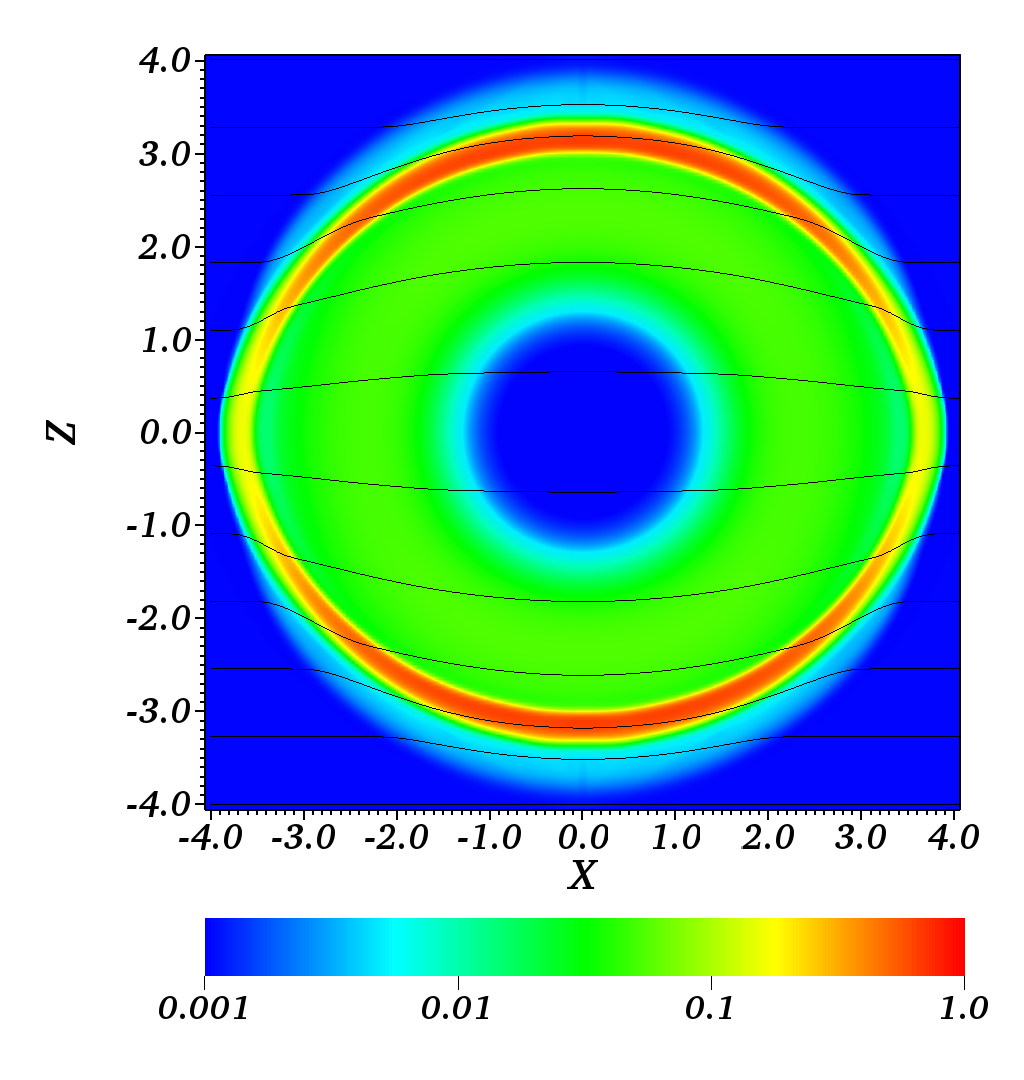}
\includegraphics[width=80mm]{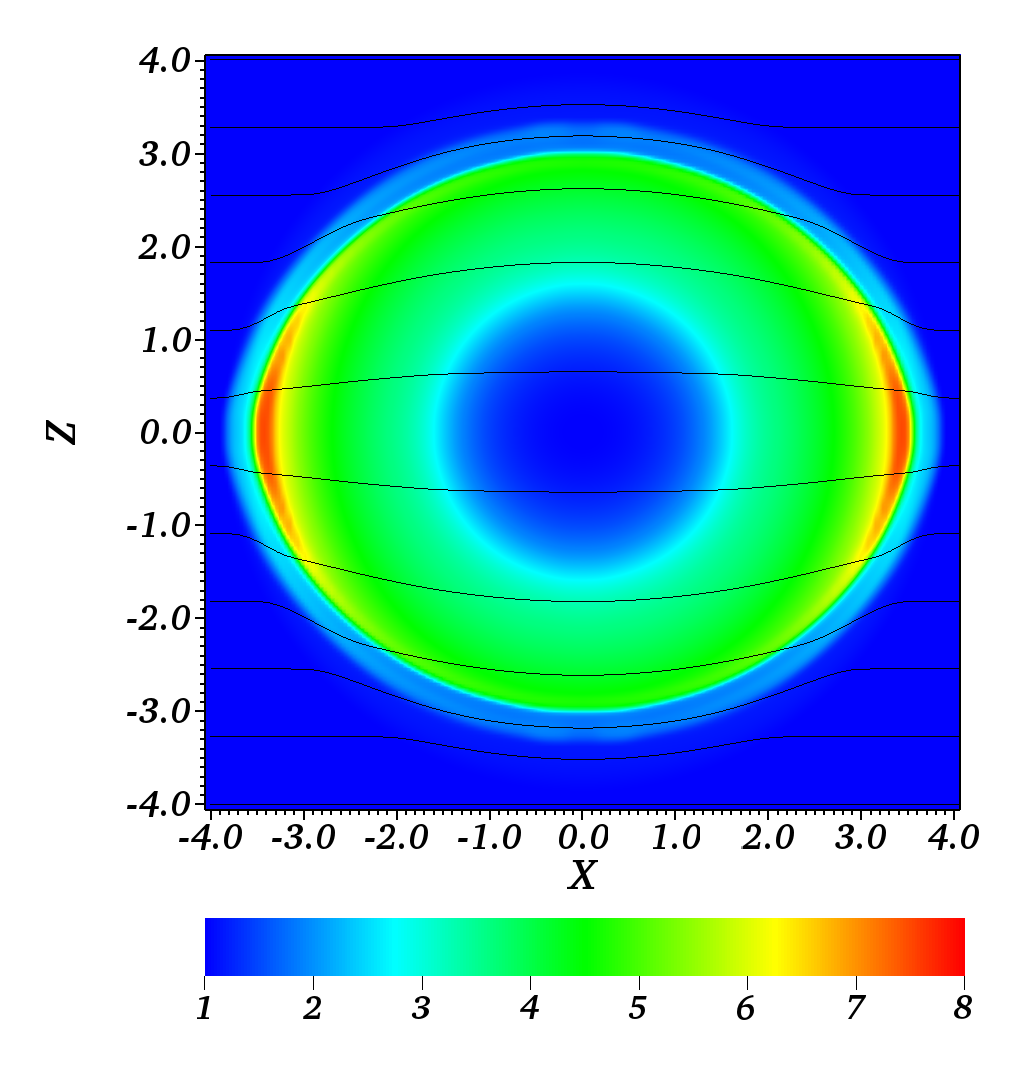}
\caption{
Spherical Blast Wave 3D test. Time $t=3$.
The left panel shows the total gas pressure ,$\log_{10} (p_+ +p_-)$, and the magnetic field lines in the 
XOZ plane.
The right panel shows the Lorentz factor $\gamma_-=\gamma_+$, the maximum value being 
$\gamma\sub{max}=7.5$.
}
\label{fig:bw3d}
\end{figure*}

The second multi-dimensional test problem deals with a spherical 
blast wave. The initial setup is the same as in the case of the cylindrical 
blast wave problem in all respect except the 
shape of the fireball, which is now a sphere of radius $r\sub{b}$. 
The computational domain is  $[-4,4]\times[-4,4]\times[-4,4]$ with a
uniform grid of $200^3$ cells. Figure~\ref{fig:bw3d} illustrates 
the solution at $t=3$. One can see that it is similar to the cylindrical 
case, with somewhat higher Lorentz factors.  There are no features which are 
grid related and hence artificial in nature.       

The results of these tests suggest that multi-fluid codes   
can be as successful at capturing the large scale dynamics of astrophysical 
flows as the single-fluid ones.

\subsection{Tearing Instability}

With the possible exception of the perpendicular shock problems, none of the above 
simulations test the inter-fluid friction term, which is analogous to resistivity. 
In the absence of exact analytic solutions of equations with non-vanishing inter-fluid 
friction term, we are forced simply to try problems 
where this term is expected to be of critical significance. 
For example, it is well known that, at least in the single 
fluid framework, it is the resistivity what drives the tearing instability 
\citep[e.g.][]{PrFo00,lyut03,ssk-tearing}. This important problem merits a separate 
comprehensive study, which we are planning to carry out in the near future. Here we 
present only the results of two runs for one particular set of parameters. 

Following \citet{ssk-tearing} we consider a 2D current sheet with the rotating  
force-free magnetic field

\beq
  \bB = B_0 [\tanh(y/l)\ort{x} + \sech(y/l)\ort{z}] \, ,
\label{b-st}
\eeq 
and a uniform static gas distribution. This configuration is perturbed via modifying
the magnetic field, $\bB \to \bB+\bb$, where

\beq
  \bb=b_0 \sin(\pi x/\lambda) \ort{y} \, .
\eeq    
The computational domain for this simulation is $[-2,+2]\times[-1,+1]$, with the periodic 
boundary conditions at the x boundaries  and the free flow boundary conditions 
at the y boundaries. 
The other parameters are $B_0=3.0$, $b_0=10^{-3} B_0$, $\lambda=2$, $l=0.1$, $n_+=n_-=1$, 
$p_+=p_-=0.1$, $\Gamma=4/3$, $\Km=\Kq=0.01$ and $\Kf=0.3$.      

\begin{figure*}
\includegraphics[width=80mm]{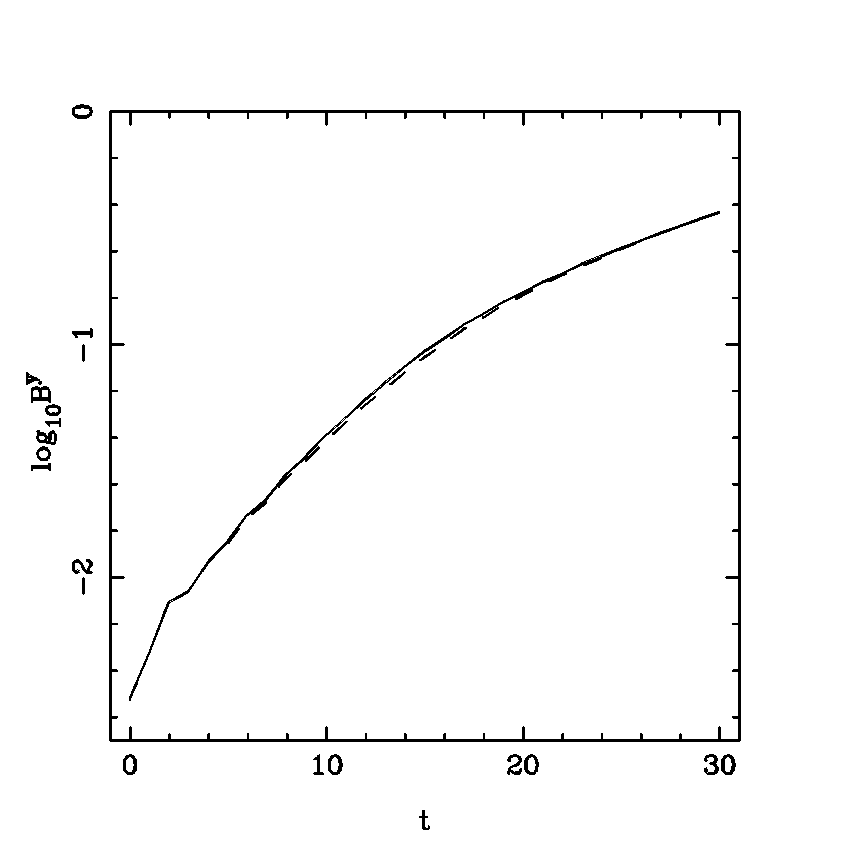}
\caption{
Convergence of the tearing instability solutions. The plots show 
$\log_{10}B^y$ as a function of time for the models with $200\times200$ 
( dashed line ) and $300\times300$ ( solid line ) cells. 
}
\label{fig:ti-growth}
\end{figure*}

Figure~\ref{fig:ti-growth} shows the evolution of the maximum value of $B^y$ 
on the grid for two models which differ only by their numerical resolution, 
one with $200\!\times\!200$ and  another with $300\!\times\!300$ cells. One can see 
the difference between the two solutions is very small, which indicates that 
the simulations have almost converged. This figure also shows that the perturbation 
grows approximately exponentially, with gradually decreasing growth rate.    
At time $t=4$ the e-folding time is $\tau_e\simeq 4.3$. 

In the approximation of resistive Magnetodynamics\footnote{Force-free Degenerate 
Electrodynamics is another name for this system of equations.} 
this problem was studied in details in \citet{ssk-tearing}. 
In order to compare our growth rate with theirs, we
need to scale parameters in the same way. Since in this problem the fluid 
motion is rather slow, $\gamma_\pm < 1.008$, the resistivity is accurately 
represented by the expression

\beq
   \eta \simeq \frac{c^2 \varkappa\sub{f}}{4\pi e^2}
\eeq
(see Eq.\ref{sigma}). The corresponding dimensionless resistivity is 
\beq
  \eta \simeq \frac{\Km \Kq}{\Kf} \simeq 3.3\times 10^{-4}\, 
\eeq
and hence the resistive time-scale of the current sheet 
$$
\tau_d = l^2/\eta \simeq 30\, .
$$ 
The dimensionless Alfv\'en speed 

$$
    v\sub{a}^2 = \frac{B^2}{B^2 + (\Km/\Kq) (w_++w_-)} \simeq 0.76\, , 
$$
leading to the Alfv\'en time-scale 
$$
    \tau_a = l/v\sub{a}\simeq 0.11  
$$ 
and the Lundquist number\footnote{In this equation, the factor $\sqrt{2}$ is introduced
following \citet{ssk-tearing}.}
$$
   \mbox{Lu} = \sqrt{2} \frac{\tau\sub{d}}{\tau\sub{a}} \sim 370\, .
$$
\citet{ssk-tearing} have shown that in the limit of Magnetodynamics, the growth 
rate of perturbations peaks at $k^* \simeq \mbox{Lu}^{-1/4} l^{-1}$  and 
the shortest e-folding time scales as 
\beq
    \tau\sub{e} \propto (\tau\sub{d} \tau\sub{a})^{1/2} \, .  
\eeq   
In fact, their simulations show that for $\mbox{Lu=140}$  the coefficient of 
proportionality is $\simeq 1.8$. 
In our test simulations $k \simeq 1.4 k^*$ and the coefficient $\simeq 2.3$.  
Thus, our results are very close to their findings. 

\begin{figure*}
\includegraphics[width=80mm]{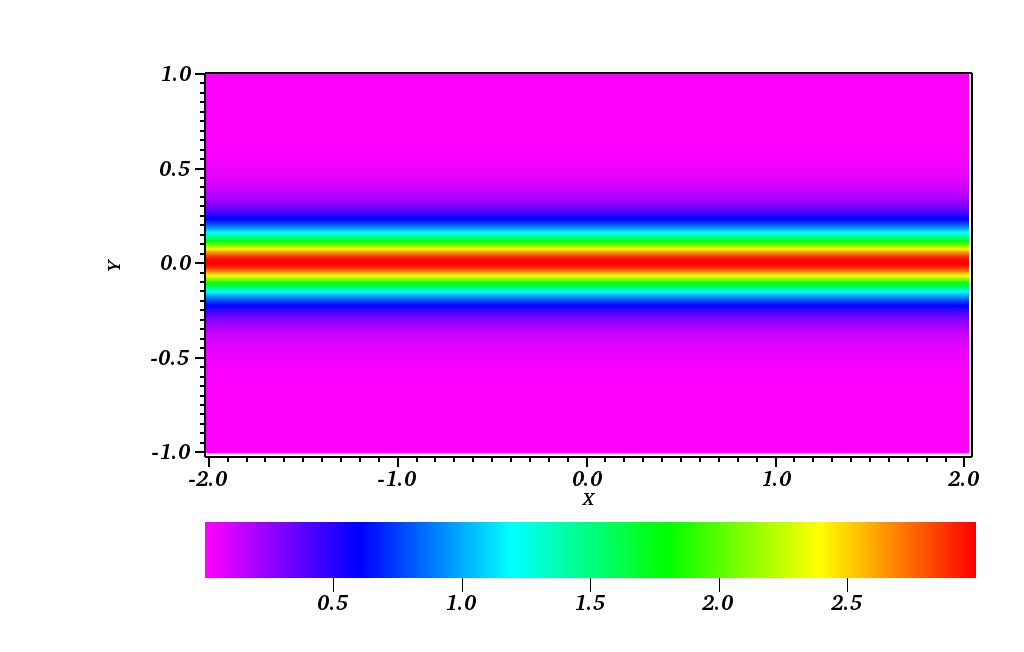}
\includegraphics[width=80mm]{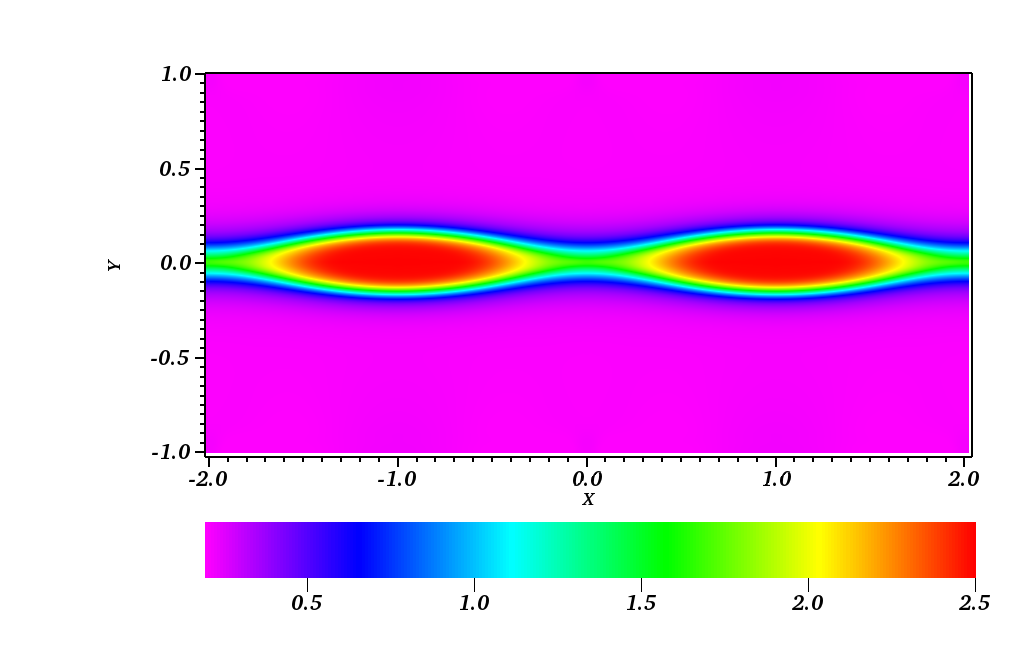}
\includegraphics[width=80mm]{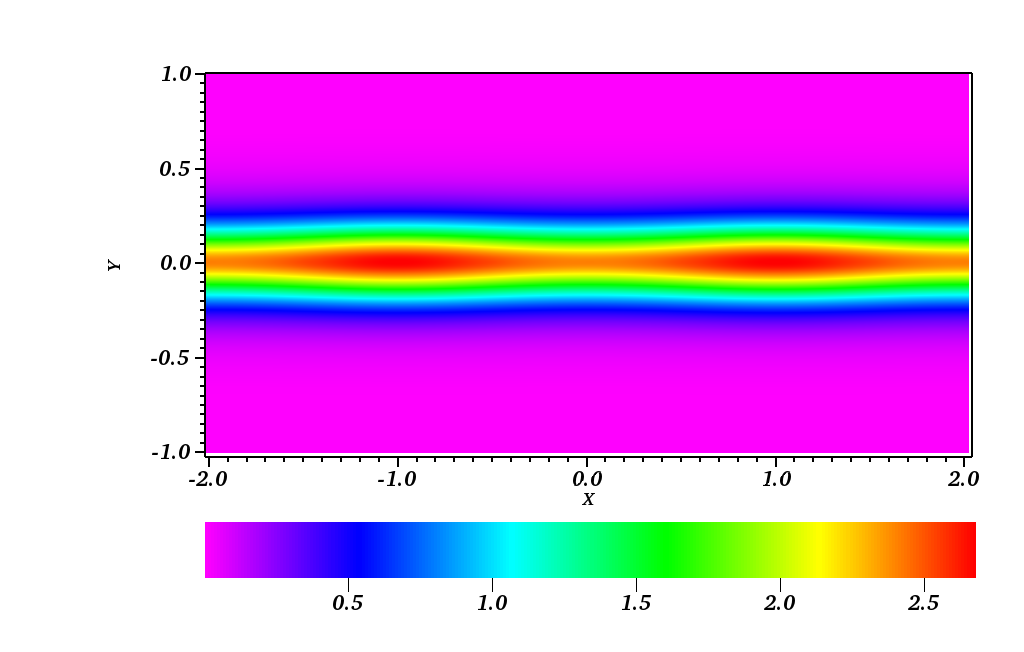}
\includegraphics[width=80mm]{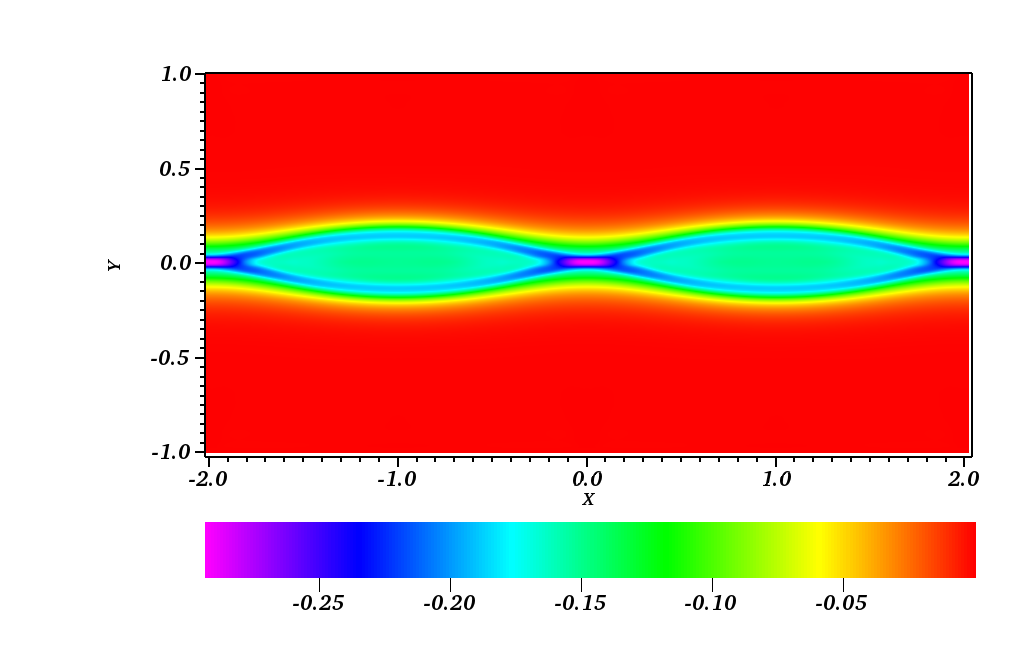}
\includegraphics[width=80mm]{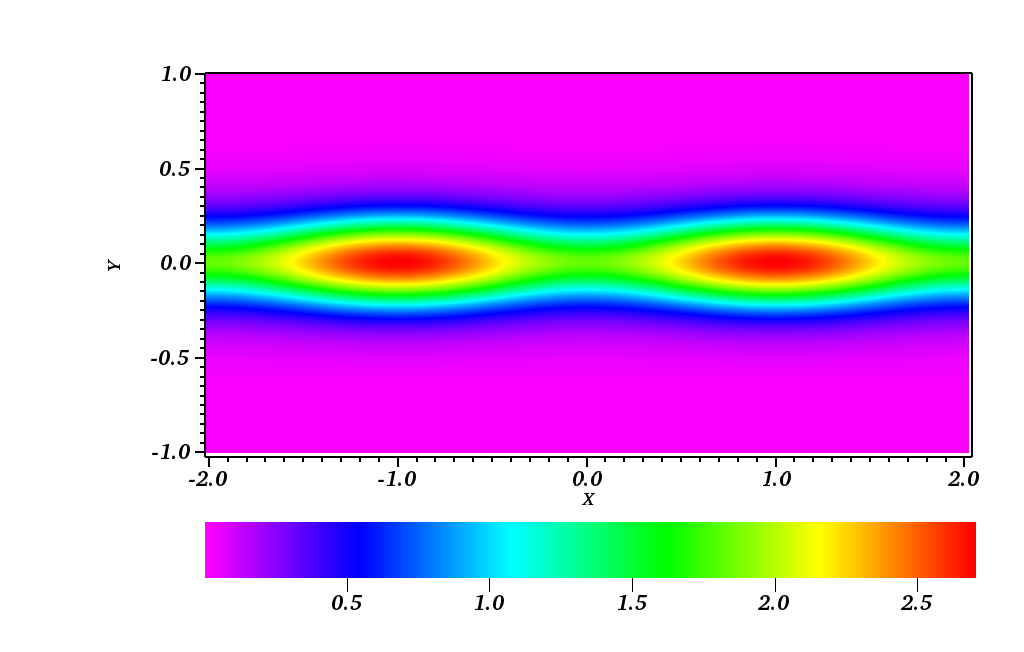}
\includegraphics[width=80mm]{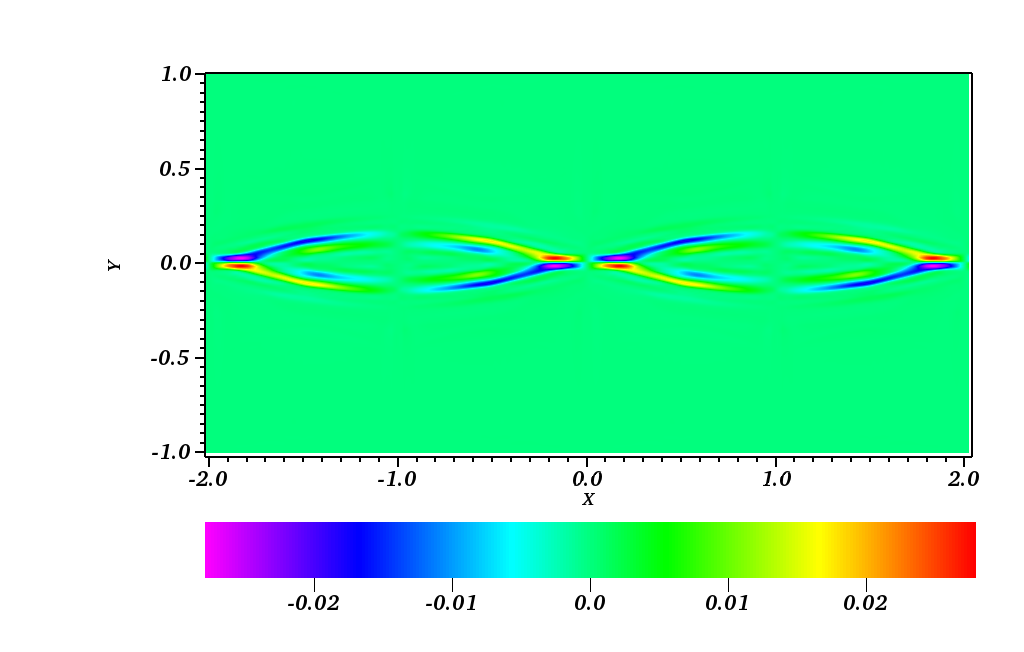}
\caption{
Tearing instability 2D test. The left panels show the $B^z$ component of the magnetic 
field at times $t=0$, 15 and 30, from top to bottom. 
The right panels show $p\sub{mat}=p_-+p_+$ ( top ), 
$j^z=n_+u^z_+-n_-u^z_-$ ( middle ) and $q=n_+\gamma_+-n_-\gamma_-$ ( bottom ) at $t=30$. 
}
\label{fig:ti}
\end{figure*}

The left panels of Figure~\ref{fig:ti} demonstrate how the instability leads to 
tearing of the current sheet during the non-linear phase. Its right panels show 
in more details the structure of the higher resolution solution by the end of the 
simulation, at $t=30$. The solution exhibits expected features, 
such as the sub-layer of strong electric current, magnetic islands etc. 

\section{Conclusion}

In this paper we considered the equations of relativistic two-fluid 
Magnetohydrodynamics for electron-positron plasma and described a 
high-order explicit finite-difference scheme for their numerical 
integration. This work was carried out as a part of the search for the 
most adequate mathematical model for the macroscopic dynamics of relativistic 
magnetized plasma, with astrophysical applications in mind. There are reasons 
to believe that this model has advantages compared to both ideal and resistive 
RMHD. Indeed, the ideal RMHD assumes $\spr{E}{B}=0$ and $E^2<B^2$ and hence allows 
magnetic dissipation only by slow shocks.  The resistive RMHD allows resistive 
magnetic dissipation but utilizes a simplistic phenomenological Ohm's law. Both 
do not capture plasma waves above the plasma frequency. The two-fluid approach 
allows more realistic Ohm's law via introduction of some microphysics into 
the system. Yet, it seems to remain oriented more towards the large-scale 
(macroscopic) dynamics. 
While other frameworks, like particle dynamics and kinetic models, 
capture the plasma microphysics much better, they are not practical for 
simulating phenomena with characteristic length scales greatly exceeding 
the gyration radius. Our scheme will help to explore the practicality of 
the two-fluid framework.  

The scheme integrates dimensionless equations which include three parameters,
$\Km$, $\Kq$, and $\Kf$, introduced by analogy with the Reynolds number.
In some applications, they may be of order unity. In others, much lower or
much higher than unity. Our numerical scheme has its limitations and will
not always be able to cope with extreme values. However, if the solution
reaches asymptotic regimes rapidly it may not be necessary to deal with
such extremes. Systematic studies of the parameter space will be required to
establish if this is the case.

The design of our numerical scheme is based on standard modern techniques 
developed for systems of hyperbolic conservation laws. In particular, 
we use third-order WENO interpolation to setup Riemann problems at the 
cell interfaces, the hyperbolic fluxes are computed using the Lax-Friedrich 
formula and the explicit time integration is carried out with third order 
TVD Runge-Kutta scheme.       
The results of test simulations presented in Sec.\ref{tests} show that the 
numerical scheme is sufficiently robust and confirm its third-order accuracy.

The test simulations revealed that in some problems including strong shocks
the two-fluid shock structure may have oscillations on the length scale below 
the skin depth. When not fully resolved, these are problematic and can lead 
to code crashing. However, they can be damped with high localized resistivity.  
A more serious limitation concerns the plasma magnetization. Similarly to
the existing explicit conservative schemes for single fluid  RMHD, 
this scheme fails when $\sigma\sub{m} \gg 1$.   
The reason for this is the very large magnetic terms compared to the gas ones in 
Eqs.\ref{total-energy},\ref{total-momentum}. Equivalently, the source terms 
in Eqs.\ref{energy},\ref{momentum} corresponding to the Lorentz force 
become very stiff in this limit. A different approach will be required to 
overcome this limitation. One possible solution is an IMEX (implicit-explicit) 
type of scheme \citep[e.g.][]{KM12}.   

Here we considered only the case of electron-positron plasma. Although, this 
is already of significant interest in many astrophysical problems, in other 
applications, protons and even neutrons need to be included. This is another 
potential direction for continuation of this work.    

The current version of our code, named JANUS, can be provided by the authors 
on request.

\section{Acknowledgments}
The authors would like to thank Anatoly Spitkovsky and Michael Gedalin for 
helpful discussions and the anonymous referee for useful comments.  
SSK acknowledges support by STFC under the standard grant ST/I001816/1.
VK, AZ and SSK acknowledge support from the Russian Ministry of Education
and Research under the state contract 14.B37.21.0915 for Federal Target-Oriented
Program. BMV acknowledges partial  support  by  RFBR  grant  12-02-01336-a.


\bibliographystyle{mn2e}
\bibliography{plasma,mypapers,lyubarsky,lyutikov,numerics,specific}
\end{document}